\documentclass[
 reprint,
 amsmath,amssymb,
 superscriptaddress,
 aps,
 prx,
 nofootinbib,
]{revtex4-2}

\usepackage{graphicx}
\usepackage{dcolumn}
\usepackage{bm}

\usepackage{physics2}
\usephysicsmodule{ab}
\usephysicsmodule{op.legacy}
\usephysicsmodule{braket}
\usephysicsmodule{nabla.legacy}
\usepackage{derivative,fixdif}

\usepackage{tensor}
\usepackage{mathtools}

\usepackage{siunitx}
\usepackage{appendix}
\usepackage[colorlinks,linkcolor=blue,urlcolor=blue, citecolor=blue]{hyperref}
\usepackage[whole]{bxcjkjatype}
\usepackage[T1]{fontenc} 
\usepackage{comment}

\newcommand{\mqty}[1]{\ab(\begin{matrix}#1\end{matrix})}
\newcommand{\dd}[1]{\d #1 \:}

\begin{document}

\title{Taming Recoil Effect in Cavity-Assisted Quantum Interconnects}

\author{Seigo Kikura}
\email{seigo.kikura@nano-qt.com}
\affiliation{Nanofiber Quantum Technologies, Inc. (NanoQT), 1-22-3 Nishiwaseda, Shinjuku-ku, Tokyo 169-0051, Japan.}
\author{Ryotaro Inoue}
\affiliation{Nanofiber Quantum Technologies, Inc. (NanoQT), 1-22-3 Nishiwaseda, Shinjuku-ku, Tokyo 169-0051, Japan.}
\author{Hayata Yamasaki}
\affiliation{Nanofiber Quantum Technologies, Inc. (NanoQT), 1-22-3 Nishiwaseda, Shinjuku-ku, Tokyo 169-0051, Japan.}
\affiliation{Department of Physics, Graduate School of Science, The Univerisity of Tokyo, 7-3-1 Hongo, Bunkyo-ku, Tokyo, 113-0033, Japan}
\author{Akihisa Goban}
\email{akihisa.goban@nano-qt.com}
\affiliation{Nanofiber Quantum Technologies, Inc. (NanoQT), 1-22-3 Nishiwaseda, Shinjuku-ku, Tokyo 169-0051, Japan.}
\author{Shinichi Sunami}
\email{shinichi.sunami@nano-qt.com}
\affiliation{Nanofiber Quantum Technologies, Inc. (NanoQT), 1-22-3 Nishiwaseda, Shinjuku-ku, Tokyo 169-0051, Japan.}
\affiliation{Clarendon Laboratory, University of Oxford, Oxford OX1 3PU, United Kingdom}

\begin{abstract}
Photon recoil is one of the fundamental limitations for high-fidelity control of trapped-atom qubits such as neutral atoms and trapped ions.
In this work, we derive an analytical model for efficiently evaluating the motion-induced infidelity in remote entanglement generation protocols.  
Our model is applicable for various photonic qubit encodings such as polarization, time bin, and frequency, and with arbitrary initial motional states, thus providing a crucial theoretical tool for realizing high-fidelity quantum networking.
For the case of tweezer-trapped neutral atoms, our results indicate that operating in the \textit{bad-cavity regime} with cavity decay rate exceeding atom-photon coupling rate, and near-ground-state cooling with motional quanta below 1, are desired to suppress the motion-induced infidelity sufficiently below the 1\% level required for efficient quantum networking.
Finite temperature effects can be mitigated efficiently by detection time filtering at the moderate cost of success probability and network speed.
These results extend the understanding of infidelity sources in remote entanglement generation protocols, establishing a concrete path towards fault-tolerant quantum networking with scalable trapped-atom qubit systems.
\end{abstract}
\maketitle

Entanglement between remote stationary qubits is the fundamental building block of quantum networking protocols \cite{Kimble2008, Wehner2018, Azuma2023} to realize multiprocessor quantum computing \cite{Covey2023, Young2022, Monroe2014, Sinclair2024, Sunami2025}, secure communication \cite{Bennett1999,Panayi2014}, long-baseline quantum sensing \cite{Gottesman2012, Khabiboulline2019}, and secure delegated quantum computing \cite{Fitzsimons2017}.
A common class of protocols for remote entanglement generation is the heralded entanglement generation (HEG) protocol consisting of atom-photon entanglement generation via an atom-state-dependent emission of photons into separate modes, such as polarization, time-bin, and frequency, and entanglement swapping via a photonic Bell-state measurement~\cite{Duan2003,Barrett2005,Beukers2024}.
The crucial metrics of these operations are the average fidelity and the rate of entanglement generation.
They have rapidly improved in recent years to reach infidelities of a few percent \cite{Ritter2012,Main2024,Saha2024} and a \qty{250}{s^{-1}} generation rate~\cite{Oreilly2024}.

However, currently, this HEG rate significantly lags behind the local gate operations of atomic qubits such as fast Raman gates \cite{Jenkins2022,Bluvstein2022} and Rydberg gates \cite{Evered2023, Peper2025}, which reach near $10^{-3}$ infidelity and MHz-order operation \cite{Bluvstein2024, Graham2022}.
Thus, current implementation may become a significant bottleneck for achieving a scalable quantum network, required for multiprocessor fault-tolerant quantum computers \cite{Sunami2025}.
To further improve physical-level HEG required for fast and resource-efficient remote fault-tolerant gate operations \cite{Ramette2024, Pattison2024}, thorough analysis and mitigation of infidelity to reach below $1\%$ \cite{Li2024, Ramette2024, Pattison2024}, as well as faster operations by multiplexed HEG in optical cavities \cite{Krutyanskiy2023, Hartung2024, Huie2021, Li2024, Sunami2025}, are crucial.

A remaining and important challenge in high-fidelity HEG protocols is the coupling of the motional and internal states of the atom and the emitted photonic states through the photon recoil.
Existing techniques to reduce recoil-induced motion-qubit coupling during single-qubit gates~\cite{Lis2023, Zhang2024} are not applicable in this case, because stochastic processes associated with photon emission and detection render deterministic mitigation impossible.
Despite its universal effect on the remote entanglement fidelity for trapped-atom quantum networking nodes, this was not analyzed in previous proposals for high-fidelity quantum networking with trapped atoms, including the comprehensive analysis of Ref.~\cite{Li2024}.
In light of the recent observation of significant motion-induced error in an experimental demonstration of time-bin-based entanglement generation with trapped ions \cite{Saha2024}, and with increasing target fidelities of remote entanglement generation for multiprocessor fault-tolerant quantum computing \cite{Li2024, Sunami2025},
it is imperative to develop an efficient theoretical toolkit covering a wide range of setups and protocols.

Here, we develop a theoretical model to perform an efficient analysis of the effect of atomic motion on the fidelity of HEG.
A key contribution of this paper is the introduction of a \textit{kick-operator} model to concisely express a time- and state-dependent photon emission including the motional effects, along with an efficient numerical evaluation protocol.
This model is applicable to free-space and cavity-enhanced emissions regardless of the choice of photonic qubit encoding, streamlining the analysis.
We demonstrate the application of our model for cavity-assisted HEG protocols which are promising for high-speed quantum networking~\cite{Young2022,Li2024,Sunami2025}, with the polarization and time-bin encodings for the photonic qubits.
We find that operating in the bad-cavity regime, where cavity decay rate exceeds atom-photon coupling rate, and the near-ground-state cooling of atoms, are desired for keeping the motion-induced infidelity below the 1\% level.
Even at a finite temperature, detection-time filtering efficeintly suppress the infidelity at the moderate cost of the success probability.
We further evaluate a time-multiplexed remote entanglement generation with time-bin encoding~\cite{Huie2021,Sunami2025} as an example of system-level rate-fidelity tradeoff evaluation.

The rest of this paper is organized as follows.
In Sec.~\ref{sec:motion-induced_infid}, we illustrate how the recoil effects induce errors in HEG protocols and introduce a kick-operator-based formalism for an efficient fidelity evaluation.
In Sec.~\ref{sec:kick_pol}, we present a derivation of the kick operators based on the analysis of atom-cavity dynamics, and evaluate the infidelity of polarization-based HEG protocols. This is then extended to the time-bin protocols in Sec.~\ref{subsec:time-bin}.
Based on the models presented, we evaluate the rate-fidelity tradeoff in time-multiplexed, time-bin-photon mediated networking operations in Sec.~\ref{sec:mux}.
Finally, we summarize our results and provide an outlook in Sec.~\ref{sec:conclusion}.

\begin{figure}[t]
    \centering
    \includegraphics[width=0.99\linewidth]{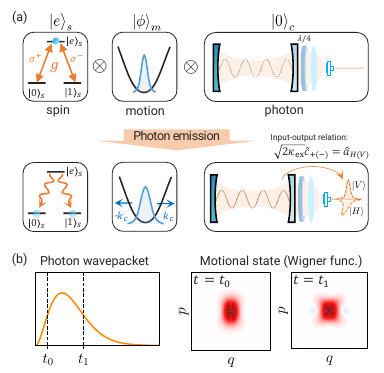}
    \caption{Motion-photon entanglement in the photon-emission process.
    We consider a trapped three-level atom, such as one in a harmonic potential, at the antinode of the cavity modes whose wavenumbers are $k_c$.
    (a) Following the preparation of the atom in $\ket{e}_s$, the atom transitions to the qubit state $\ket{0}_s(\ket{1}_s)$ while generating a $\sigma^{+(-)}$ photon inside the cavity, which leaks through a coupling mirror and passes through the quarter-wave plate.
    Through this process, the internal, motional, and photonic states become entangled. 
    (b) Stochastic detection times lead to different motional states from time-dependent recoil kicks, resulting in infidelity of the generated entanglement that we model and analyze in detail in this work.
    }
    \label{fig:photon_gen}
\end{figure}

\section{Spin-motion-photon coupling and entanglement fidelity}
\label{sec:motion-induced_infid}

To illustrate how spin-motion-photon coupling arises in the quantum networking protocols and leads to infidelities, we consider a simple case where a networking node consists of an atom placed in a polarization-degenerate optical cavity as shown in Fig.~\ref{fig:photon_gen}.
As a simplified setup for analyzing the HEG protocol, we assume that the atom has internal qubit states $\ket{0}_s, \ket{1}_s$ as well as initial and excited states $\ket{\text{init}}_s$, $\ket{e}_s$, and the transition $\ket{0}_s\leftrightarrow \ket{e}_s$($\ket{1}_s\leftrightarrow \ket{e}_s$) is resonantly coupled to the $\sigma^{+(-)}$-polarized cavity mode.
Following the excitation $\ket{\text{init}}_s\to\ket{e}_s$, the atom decays from $\ket{e}_s$ to the qubit states $\ket{0}_s$ and $\ket{1}_s$ with equal probability, respectively generating $\sigma^{+}$- and $\sigma^{-}$-polarized photons inside the cavity. 
The photons leave the cavity through the coupling mirror and pass through the quarter-wave plate, resulting in the spin-photon entanglement state $(\ket{0}_s\ket{H}+\ket{1}_s\ket{V})/\sqrt{2}$ where $\ket{H(V)}$ represents the $H(V)$-polarized itinerant photon.
Including the nonzero spread of the photon emission times, the entangled state is expressed by
\begin{equation} \label{eq:ideal_atom_photon_entanglement}
    \frac{1}{\sqrt{2}}\left( \ket{0}_s \int \dd{t} f(t) \hat{a}_{H}^\dagger(t) + \ket{1}_s \int \dd{t} f(t) \hat{a}_{V}^\dagger(t)\right) \ket{\text{vac}},
\end{equation}
where $\ket{\text{vac}}$ represents the vacuum state of the propagating mode, $\hat{a}_{H,V}(t)$ represent instantaneous annihilation operators for the two orthogonal polarizations, and $f(t)$ is the envelope function for the photon wavepacket~\cite{Reiserer2015,Krutyanskiy2023,Li2024}. 

However, in the case of trapped atoms, such as neutral atoms in optical tweezers and ions in electromagnetic traps, the excitation and photon-emission induced recoil result in a disturbance of the motional state.
For simplicity, we consider that the initial spin-motion system is a pure product state $\ket{\text{init}}_s \otimes \ket{\phi}_m$, where $\ket{\phi}_m$ is the external (motional) atomic state---the extension to a mixed state is straightforward and will be discussed later in this section.
The excitation and photon recoil result in a state with spin-motion-photon entanglement of the form
\begin{equation}
    \label{eq:atom-photon_entanglement}
    \begin{aligned} 
    \ket{\Phi} = \frac{1}{\sqrt{2}} \Big( & \ket{0}_s \int \dd{t} \hat{\mathcal{K}}_0(t) \hat{a}_{H}^\dagger(t) \\
    &+ \ket{1}_s \int \dd{t} \hat{\mathcal{K}}_1(t) \hat{a}_{V}^\dagger(t) \Big) \ket{\phi}_m \otimes \ket{\text{vac}},
    \end{aligned}
\end{equation}
where we have introduced a \textit{kick operator} $\hat{\mathcal{K}}_{i}(t)$ acting on the motional state, incorporating the recoil effect of the excitation and photon emission.
This operator reduces to an identity operator with prefactor $f(t)$ in the limit of large atomic mass or tight trap, recovering the expression of Eq.~\eqref{eq:ideal_atom_photon_entanglement}.
This recoil-induced entanglement among the three (spin, motional, and photonic) degrees of freedom eventually reduces the fidelity of the generated atom-atom entanglement in a HEG protocol, as described in the following.

\begin{figure}
    \centering
    \includegraphics[width=\linewidth]{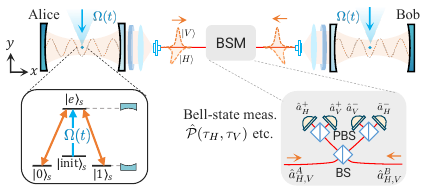}
    \caption{HEG process with polarization encoding. Alice and Bob generate the spin-photon entanglement, and the photons are sent to the Bell-state measurement (BSM) apparatus consisting of a balanced beam splitter (BS), polarizing beam splitters (PBS), and photon detectors.
    }
    \label{fig:pol_heg}
\end{figure}
To establish a remote atom-atom entanglement, the two parties Alice and Bob respectively generate the spin-motion-photon entanglement $\ket{\Phi}^A$ and $\ket{\Phi}^B$ in Eq.~\eqref{eq:atom-photon_entanglement}, where we add the superscripts $q\in\{A,B\}$ to all operators and states except $\ket{\text{vac}}$ to distinguish the two parties.
Following the generation of spin-photon entanglement via photon emission, the HEG protocol completes with a measurement of photons in the Bell basis, i.e., Bell-state measurement (BSM), at the middle node, comprising the interference of photons at a nonpolarizing beamsplitter and a pair of single-photon detectors, as illustrated in Fig.~\ref{fig:pol_heg}. 
A balanced beamsplitter interferes the modes $\hat{a}_{H(V)}^A $ and $\hat{a}_{H(V)}^B$, resulting in output modes $\hat{a}^\pm_{H(V)} = (\hat{a}_{H(V)}^A \pm \hat{a}_{H(V)}^B)/\sqrt{2}$.
The detection of photons at the output of the beamsplitters with the desired patterns projects the remote atom pair to one of two Bell states $\ket{\Psi^{\pm}} = (\ket{01}_s \pm \ket{10}_s)/\sqrt{2}$, where $\ket{ij}_s = \ket{i}_s^A\ket{j}_s^B$, in the case of no recoil-induced errors~\cite{Simon2003}. 
For example, the case of single-photon detection at the modes $\hat{a}^+_{H}$ and $\hat{a}^+_{V}$ corresponds to the generation of $\ket{\Psi^+}$.
Here, reflecting the nonzero spread of the photon emission times, the detection times $\tau_{H,V}$ for the $H$ and $V$ modes are stochastically determines.
Formally, such a detection event can be expressed by a measurement operator $\hat{\mathcal{P}}(\tau_H,\tau_V) = \hat{a}^+_H(\tau_H)\hat{a}^+_V(\tau_V)$.
The final state of the successful HEG process, conditioned on the measurement pattern as discussed above with detection times $(\tau_H, \tau_V)$, is obtained by the post-emission state $\ket{\Phi}^A\ket{\Phi}^B$ and the measurement operator as follows:
\begin{equation*}
    \label{eq:post_measurement_state}
    \begin{aligned}
        &\bra{\text{vac}} \hat{\mathcal{P}}(\tau_H,\tau_V) \ket{\Phi}^A\ket{\Phi}^B  \\
        \propto& \ab[\ket{01}_s \hat{\mathcal{K}}_0^{A}(\tau_H)\hat{\mathcal{K}}_1^{B}(\tau_V) + \ket{10}_s \hat{\mathcal{K}}_1^{A}(\tau_V)\hat{\mathcal{K}}_0^{B}(\tau_H)]\ket{\phi}_m^{A}\ket{\phi}_m^{B},
    \end{aligned}
\end{equation*}
with proper normalization.
Tracing out motional states gives the expression of the spin-spin entanglement $\hat{\rho}(\tau_H,\tau_V)$. 
Assuming here identical systems of Alice and Bob\footnote{We analyze the case of differing system parameters between Alice and Bob in Appendix~\ref{ap:infidelity_from_parameter_difference}.}, i.e., $\hat{\mathcal{K}}_{0(1)}^{A} = \hat{\mathcal{K}}_{0(1)}^{B} (= \hat{\mathcal{K}}_{0(1)})$ and $\ket{\phi}_m^{A} = \ket{\phi}_m^{B}$, we find
\begin{equation}
    \label{eq:final_state_bell_basis}
    \begin{aligned}
         \hat{\rho}(\tau_H,\tau_V) =& \frac{1+J(\tau_H,\tau_V)}{2}\ketbra*{\Psi^{+}}{\Psi^{+}}\\
         &+ \frac{1-J(\tau_H,\tau_V)}{2}\ketbra*{\Psi^{-}}{\Psi^{-}},
    \end{aligned}
\end{equation}
with infidelity 
\begin{equation}
    \label{eq:fidelity_expression}
    \epsilon(\tau_H,\tau_V) = \frac{1-J(\tau_H,\tau_V)}{2},
\end{equation}
to the desired Bell state $\ket{\Psi^+}$, consisting only of the $Z$-type (phase) error.
The parameter $J(\tau_H,\tau_V)$ is obtained from the kick operators.
Expressing a motional-mode initial state more generally with a density operator (again for simplicity assuming that Alice and Bob start with the same state: $\hat{\rho}_m^A = \hat{\rho}_m^B \eqqcolon \hat{\rho}_m$) gives (Appendix \ref{ap:derivation_of_infid} provides a detailed derivation)
\begin{equation} \label{eq:general_J}
    J(\tau_H,\tau_V) = \frac{\Tr[\hat{\mathcal{K}}_{0}(\tau_H)\hat{\rho}_m\hat{\mathcal{K}}^\dagger_{1}(\tau_V)]\Tr[\hat{\mathcal{K}}_{1}(\tau_V)\hat{\rho}_m \hat{\mathcal{K}}^\dagger_{0}(\tau_H)]}{\Tr[\hat{\mathcal{K}}_{0}(\tau_H)\hat{\rho}_m \hat{\mathcal{K}}^\dagger_{0}(\tau_H)]\Tr[\hat{\mathcal{K}}_{1}(\tau_V)\hat{\rho}_m \hat{\mathcal{K}}^\dagger_{1}(\tau_V)]},
\end{equation}
which can be computed from the expressions for the initial motional state $\hat{\rho}_m$ and the protocol-specific kick operator $\hat{\mathcal{K}}_i(t)$.
Furthermore, the probability density of the detection is
\begin{equation}
    \begin{aligned}
        &P_\text{D}(\tau_H,\tau_V) \\
        =& \Tr[\hat{\mathcal{K}}_{0}(\tau_H)\hat{\rho}_m\hat{\mathcal{K}}^\dagger_{0}(\tau_H)]\Tr[\hat{\mathcal{K}}_{1}(\tau_V)\hat{\rho}_m\hat{\mathcal{K}}^\dagger_{1}(\tau_V)]/8.
    \end{aligned}
\end{equation}
We note that the detection pattern $\{\hat{a}_H^-(\tau_H),\hat{a}_V^-(\tau_V)\}$ also projects the atomic state to $\ket{\Psi^+}$, and $\{\hat{a}_H^+(\tau_H), \hat{a}_V^-(\tau_V)\}$ and $\{\hat{a}_H^-(\tau_H), \hat{a}_V^+(\tau_V)\}$ to $\ket{\Psi^-}$, with the same expression for infidelity $\epsilon(\tau_H,\tau_V)$ and probability density $P_\text{D}(\tau_H,\tau_V)$; overall success probability density of the HEG is thus $4P_\text{D}(\tau_H,\tau_V)$.
The above discussion can be generalized to other photonic qubit encodings such as time bin and frequency; see Sec.~\ref{subsec:time-bin} and Appendix~\ref{ap:derivation_of_infid}.

In the following, we will analyze the concrete expressions for the recoil-induced infidelity by constructing kick operators for polarization- and time-bin-based protocols, starting from the microscopic Hamiltonian evolution.

\section{derivation of the kick operator: polarization encoding}
\label{sec:kick_pol}
We now turn our attention to the microscopic Hamiltonian dynamics to obtain the kick operators.
The atom-cavity system is characterized by the atom-photon coupling rate $g$, the coupling rate between the intra-cavity mode and the (desired) external propagating mode $\kappa_{\text{ex}}$, and the other intra-cavity photon decay rate $\kappa_{\text{in}}$, where the total cavity decay rate is given by $\kappa = \kappa_\text{ex}+\kappa_\text{in}$.
The atom has a spontaneous decay from $\ket{e}_s$ at the polarization decay rate $\gamma$ and is trapped in a state-insensitive harmonic potential with trap angular frequency $\omega_\mu$ for trap axis $\mu=\{x,y,z\}$.
As illustrated in Fig.~\ref{fig:pol_heg}, we consider the excitation pulse to be applied along a direction (denoted $y$) that is orthogonal to the cavity mode (along $x$). 
In the following, we assume that the excitation to $\ket{e}_s$ is driven by a fast $\pi$ pulse with Rabi frequency $\Omega(t)$ being sufficiently larger than the atomic decay and cavity-response rate: $\Omega(t) \gg \gamma$ and $\Omega(t) \gg \min(\kappa, g^2/\kappa)$.
The former condition is crucial to avoid the reexcitation-induced infidelity~\cite{Krutyanskiy2023,Li2024,Tanji2024,Kikura2024}, whereas the latter allows us to simplify the atom-motion dynamics; under this realistic condition, the excitation and emission processes can be treated as a sequence of two independent evolutions along orthogonal axes (see Appendix~\ref{ap:separation_excitation_photon_gen}).
The excitation-induced recoil effect is expressed as a unitary operator $\hat{R}_y$ acting on the motional degree of freedom along the $y$ axis. 
For $\Omega(t) \gg \omega_y$, this becomes the displacement operator $e^{i\eta_y (\hat{b}_y^\dagger + \hat{b}_y)}$ where $\eta_y$ is the Lamb-Dicke (LD) parameter of the excitation~\cite{Poyatos1996}.
The Hamiltonian in the rotating frame of the cavity frequency is given by
\begin{equation} \label{eq:H_w/o_rotaing_frame_pol}
    \begin{aligned}
        \hat{\bar{H}}_\text{p} =& \sum_\mu \hbar\omega_\mu \hat{b}_\mu^\dagger\hat{b}_\mu \\
        &+ \hbar g \cos(k_c\hat{q}_x) [(\hat{c}_+   \ketbra{e}[_s]{0} + \hat{c}_- \ketbra{e}[_s]{1}) + \text{H.c.}],
    \end{aligned}
\end{equation}
where $\hat{c}_{+(-)}$ is the annihilation operator of the cavity mode corresponding to $\sigma^{+(-)}$ polarization, $\hat{b}_\mu$ is the annihilation operator of the motional mode along axis $\mu$, $\hat{q}_x = \sqrt{\hbar/2m\omega_x}(\hat{b}_x+\hat{b}_x^\dagger)$ is the atomic position operator along $x$ axis for the atomic mass $m$, and $k_c$ is the wavenumber of the cavity fields.
Here, $g \cos(k_c\hat{q}_x)$ represents the position-dependent coupling strength (operator).
Further, in the rotating frame of relevant trap frequencies, the Hamiltonian is
\begin{equation} \label{eq:pol_Hamiltonian}
    \begin{aligned}
        \hat{H}_\text{p}(t) =&  \hbar g \cos[\eta_x(\hat{b}_xe^{-i\omega_x t}+\hat{b}_x^\dagger e^{i\omega_x t})]\\
        &\times  [(\hat{c}_+   \ketbra{e}[_s]{0} + \hat{c}_- \ketbra{e}[_s]{1}) + \text{H.c.}],
    \end{aligned}
\end{equation}
where $\eta_x = k_c\sqrt{\hbar/2m\omega_x}$ is the LD parameter along the cavity axis ($x$ axis).

To derive the spin-motion-photon state, we consider the dynamics under the condition of no atomic and cavity decays, which is described by the non-unitary evolution operator as follows:
\begin{equation}\label{eq:hamiltonian_dynamics}
    \hat{O}_{\mathcal{H}_\text{p}}(t;t_0) = \mathcal{T}\ab[\exp\ab(-\frac{i}{\hbar}\int_{t_0}^{t} \dd{t^\prime} \hat{\mathcal{H}}_\text{p}(t^\prime))],
\end{equation}
where $\hat{\mathcal{H}}_\text{p}(t) \coloneqq \hat{H}_\text{p}(t) - i\hbar[\gamma \ketbra{e}[_s]{e}+ \kappa(\hat{c}_+^\dagger\hat{c}_+ + \hat{c}_-^\dagger\hat{c}_-)]$ is the non-Hermitian Hamiltonian, and $\mathcal{T}[\cdot]$ is the time-ordering superoperator.
This allows us to obtain the propagating mode dynamics through the following input-output relations:
\begin{equation}
    \begin{aligned}
        \hat{K}_{+}(t)\hat{a}^\dagger_{H}&(t)\ket{0}_s\ket{0}_c\ket{\text{vac}}/\sqrt{2} \\
        =& \sqrt{2\kappa_\text{ex}}\hat{c}_{+} \hat{O}_{\mathcal{H}_\text{p}}(t;0) \ket{e}_s\ket{0}_c\ket{\text{vac}}, \\
        \hat{K}_{-}(t)\hat{a}^\dagger_{V}&(t)\ket{1}_s\ket{0}_c\ket{\text{vac}}/\sqrt{2} \\
        =& \sqrt{2\kappa_\text{ex}}\hat{c}_{-} \hat{O}_{\mathcal{H}_\text{p}}(t;0) \ket{e}_s\ket{0}_c\ket{\text{vac}},
    \end{aligned}
\end{equation}
where $\ket{0}_c$ is the vacuum state of two cavity modes, and the recoil effects in the Hamiltonian dynamics~\eqref{eq:hamiltonian_dynamics} are represented by operators $\hat{K}_{\pm}(t)$ acting on the motional degrees of freedom.
Concise expressions can be obtained by rearrangements:
\begin{equation}
    \begin{aligned}
        \hat{K}_{+}(t) =& 2\sqrt{\kappa_\text{ex}} \tensor[_s]{\bra{0}_c\bra{0}}{} \hat{c}_{+}\hat{O}_{\mathcal{H}_\text{p}}(t;0)\ket{e}_s\ket{0}_c, \\
        \hat{K}_{-}(t) =& 2\sqrt{\kappa_\text{ex}} \tensor[_s]{\bra{1}_c\bra{0}}{} \hat{c}_{-}\hat{O}_{\mathcal{H}_\text{p}}(t;0)\ket{e}_s\ket{0}_c.
    \end{aligned}
\end{equation}
The overall recoil effects induced by the protocol are the product of the excitation-laser-induced recoil $\hat{R}_y$ and the emission-based kicks $\hat{K}_\pm(t)$:
\begin{equation}
    \hat{\mathcal{K}}_{0}(t) = \hat{K}_{+}(t) \hat{R}_y, \quad \hat{\mathcal{K}}_{1}(t) = \hat{K}_{-}(t) \hat{R}_y.
\end{equation}
See also Appendix~\ref{ap:derivation_of_kick_operators} for the detailed step-by-step derivation.
Substituting kick operators into Eq.~\eqref{eq:general_J} gives
$\epsilon(\tau_H, \tau_V)= [1-J_x(\tau_H,\tau_V)]/2$, where  
\begin{equation} \label{eq:J_x}
    \begin{aligned}
        J_x(&\tau_H,\tau_V) \\
        =& \frac{\Tr[\hat{K}_{+}(\tau_H)\hat{\rho}_{m,x}\hat{K}^\dagger_{-}(\tau_V)]\Tr[\hat{K}_{-}(\tau_V)\hat{\rho}_{m,x}\hat{K}^\dagger_{+}(\tau_H)]}{\Tr[\hat{K}_{+}(\tau_H)\hat{\rho}_{m,x} \hat{K}^\dagger_{+}(\tau_H)]\Tr[\hat{K}_{-}(\tau_V)\hat{\rho}_{m,x} \hat{K}^\dagger_{-}(\tau_V)]},
    \end{aligned}
\end{equation}
and $\hat{\rho}_{m,x}$ is the initial motional state along $x$ axis.
The excitation-laser-induced recoil $\hat{R}_y$ completely factors out from the expression since they act the same way for two qubit states.
Equation~\eqref{eq:J_x} only depends on the $x$-axis motional degree of freedom, thanks to the well-defined single emission mode along $x$ axis in the cavity-assisted protocols.
This is in contrast to HEG protocols with free-space photon emission without a cavity, where the photon emits at a random angle and thus causes motional effects along all three directions~\cite{Saha2024}.
The kick operators $\hat{K}_\pm(t)$ can be efficiently evaluated with standard numerical routines even in the presence of the time-ordering superoperator, as described in Appendix~\ref{ap:numerical_calculation_method_of_K(t)}.

\begin{figure}
    \centering
    \includegraphics[width=\linewidth]{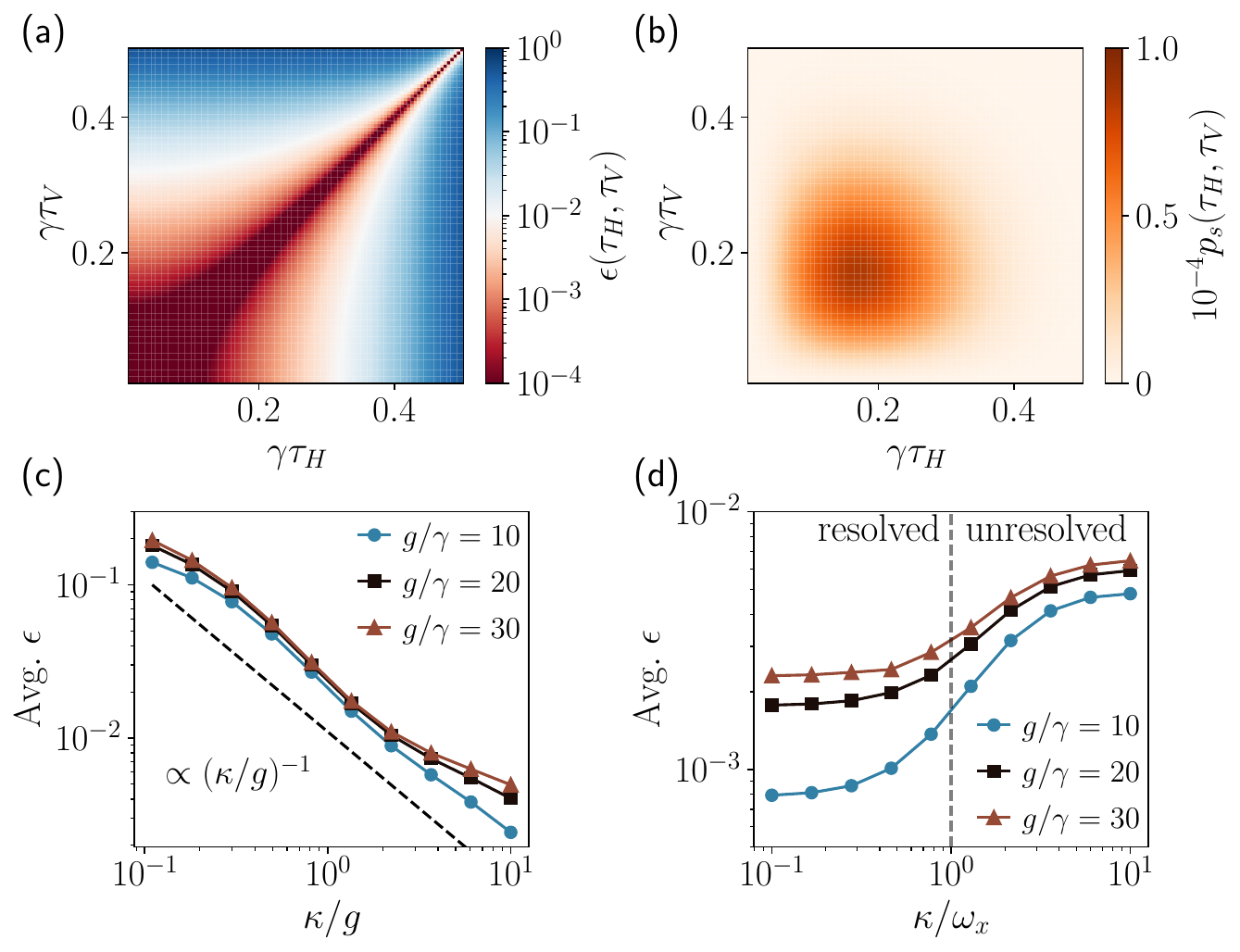}
    \caption{Motion-induced infidelity in the polarization-encoding HEG protocol. 
    (a) Infidelity and (b) success probability of the HEG protocol as a function of detection times $(\tau_H,\tau_V)$, using parameters $(g, \kappa_\mathrm{ex}^\mathrm{opt}, \kappa_{\text{in}},\omega_x) = (5.0, 5.1, 1.0, 0.10)\gamma$, and $(\eta_x, \bar{n}_x) = (0.20, 1.0)$. 
    Here, $p_s(\tau_H, \tau_V) = 4P_D(\tau_H, \tau_V)\Delta \tau^2$ with the time step $\Delta\tau = 0.005/\gamma$.
    (c) Dependence of the average infidelity on the ratio $\kappa/g$, using $(\kappa_\mathrm{in},\omega_x) = (1, 0.1)\gamma$, and $(\eta_x, \bar{n}_x)=(0.2, 1)$.
    The infidelity is observed to scale as $(\kappa/g)^{-1}$, as shown by a black dashed line, indicating that $\kappa \gg g$ is required to achieve sub-percent level infidelity.  
    (d) Dependence on the ratio of $\kappa/\omega_x$ at $\kappa/g=10$. 
    The infidelities exhibit different characteristics depending on the values of $g$, transitioning between the motional sideband-resolved regime ($\kappa/\omega_x < 1$), which yields lower infidelity, and the unresolved regime ($\kappa/\omega_x > 1$) with higher errors; see text.
    }\label{fig:pol_infid_prob_results}
\end{figure}
We now investigate the motion-induced infidelity in the relevant parameter regime for a trapped atom in a cavity, where $\hat{\rho}_{m, x}$ is a thermal state with the mean phonon number $\bar{n}_x$, examining both its magnitude and the efficacy of error-suppression strategies.
In Fig.~\ref{fig:pol_infid_prob_results}(a), we illustrate the motion-induced infidelity as a function of detection times $(\tau_H, \tau_V)$, together with detection probabilities in Fig.~\ref{fig:pol_infid_prob_results}(b), using the parameters $(g, \kappa_\mathrm{ex}^\mathrm{opt}, \kappa_\mathrm{in},\omega_x)=(5.0,5.1,1.0,0.10)\gamma$.
Here, we choose the external coupling rate as $\kappa_\mathrm{ex}^\mathrm{opt}=\kappa_\mathrm{in}\sqrt{2C_\mathrm{in}+1}$ with the internal cooperativity $C_\mathrm{in}= g^2/(2\kappa_\mathrm{in}\gamma)$, which maximizes the upper bound of photon generation probability~\cite{Goto2019}.
For large detection-time difference $|\tau_H-\tau_V|$, the infidelity significantly increases, reaching the 10~\% level.

To identify the operational conditions for achieving high-fidelity HEG, we evaluate the dependence of the pulse-averaged infidelity on the ratio $\kappa/g$ and $\kappa/\omega_x$, as shown in Fig.~\ref{fig:pol_infid_prob_results}(c) and (d), respectively.
In the strong coupling regime ($g>\kappa$), the photon generated inside the cavity can be absorbed by the atom before leaving the cavity and then regenerated.
This preocess leads to multiple photon-induced recoil kicks and results in large infidelity, as shown in Fig.~\ref{fig:pol_infid_prob_results}(c). 
To suppress this effect, the system should be operated in the bad cavity regime ($\kappa > g$), ensuring that the photon leaves the cavity before being reabsorbed. 
For small $\eta_x$, this confines the changes in motional quanta to $\Delta n_x=0, \pm 2$ because of $\cos[\eta_x(\hat{b}_x + \hat{b}_x^\dagger)] \approx 1 -\eta_x^2(\hat{b}_x + \hat{b}_x^\dagger)^2/2$, such that larger $\kappa/g$ results in smaller motion-induced infidelity with a common scaling $(\kappa/g)^{-1}$.

Another crucial consideration is whether the motional sideband is resolved ($\kappa/\omega_x < 1$) or unresolved ($\kappa/\omega_x > 1$) from the cavity mode -- in the resolved regime, the cavity enhances photon emission only on the carrier transition $\Delta n_x = 0$, thus suppressing changes in the motional quanta through the photon emission and resulting in improved fidelity (Fig.~\ref{fig:pol_infid_prob_results}(d)).
This is particularly relevant for trapped ion platforms where the trap frequency is on the order of MHz, typically comparable to or larger than $g$, while tweezer-trapped neutral atoms mostly operate in the unresolved regime.

\begin{figure}
    \centering
    \includegraphics[width=\linewidth]{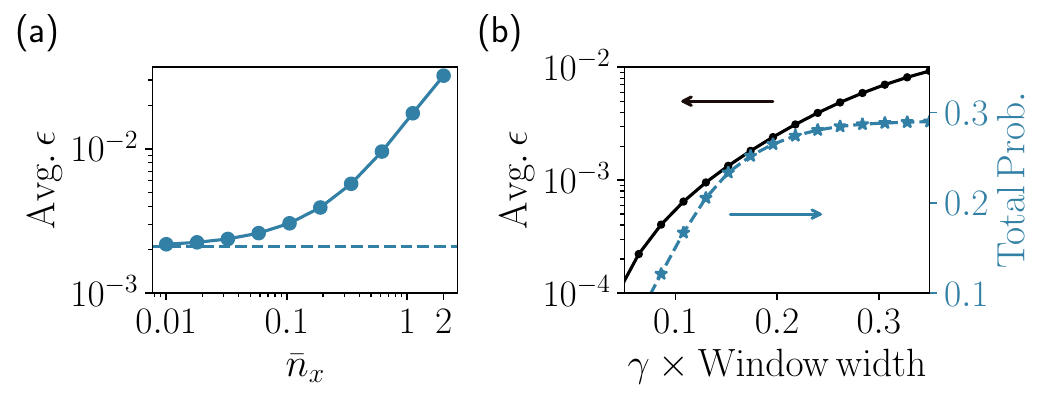}
    \caption{Mitigating the recoil-induced infidelity.
    (a) Dependence of the infidelity on the mean phonon number $\bar{n}_x$.
    Cooling to the near-ground state $\bar{n}_x < 1$ allows suppression of $\epsilon$.
    The dotted lines represent the averaged infidelity at $\bar{n}_x = 0$.
    (b) Efficient error suppression by the detection time filtering for finite-temperature atoms. Infidelity $\epsilon$ (black, left axis) and the HEG success probability (blue, right axis) change significantly by detection-time filtering, where the window is centered at the peak of temporal photon pulse (Fig.~\ref{fig:pol_infid_prob_results}(b)).
    The parameters are $(g, \kappa_\text{ex}, \kappa_\text{in}, \omega) = (10, 11, 1, 0.1)\gamma$ and $\bar{n}_x = 1$.
    }
    \label{fig:window_temperature}
\end{figure}

Operationally, a high entanglement generation rate requires $\kappa \sim g$, but the rate rapidly decreases for large $\kappa/g$~\cite{Sunami2025},
thus limiting the achievable fidelity while maintaining reasonable networking speed.
Atom cooling close to ground state is effective for achieving high fidelity (Fig.~\ref{fig:window_temperature}(a)), however, requires regular cooling sequences with large time and operational cost.
Even for finite temperature samples, detection-time filtering is an efficient strategy for improving the average fidelity of generated entanglement; discarding cases with large errors near the tail of the temporal photon distribution, as shown in Fig.~\ref{fig:window_temperature}(b).
Other filtering strategies can be devised, for example, by using the contour of a specific infidelity level in the detection-time-dependent infidelity plot (Fig.~\ref{fig:pol_infid_prob_results}(a)) as a custom window to bound motion-induced errors.

Our model can also be extended to incorporate more realistic scenarios, including mismatches in the experimental parameters of two different atom-cavity systems, by using different kick operators for the two atoms, as discussed in Appendix~\ref{ap:infidelity_from_parameter_difference}.
We have further applied our model to a running-wave cavity where photon generation occurs unidirectionally and the changes in motional quanta become $\Delta_x = 0, \pm 1$, resulting in a modified scaling (Appendix \ref{ap:running_wave_cavity}).

\section{Time-bin encoding} \label{subsec:time-bin}
Time-bin encoding of photonic qubits is a promising approach for quantum networking, offering robustness against optical-fiber-induced polarization and phase fluctuations~\cite{Knaut2024, Huie2021}, compatibility with nondegenerate polarization-mode cavity systems such as nanophotonic resonators~\cite{Kato2015, Tamara2021, Gonzalez2024}, and scalability for extending to the higher-dimensional qudit-based networking \cite{Zheng2023}.
As schematically illustrated in Fig.~\ref{fig:time_bin_schematic}(a), however, the state-dependent photon emission requires additional steps, which must be carefully considered to accurately evaluate and mitigate motion-induced infidelity.
The initial state is a superposition of the qubit states: $\ket{\text{init}}_s = (\ket{0}_s+\ket{1}_s)/\sqrt{2}$, from which selective excitation from $\ket{1}_s$ to the excited state is performed, followed by cavity-enhanced photon emission, completing the generation of the `early' time-bin photon.
Typically, the `late' time-bin photon generation follows the state swap (Pauli $X$ gate) with separation in time $T$ between the two excitations, resulting in an atom-motion-photon entangled state of the form Eq.~\eqref{eq:atom-photon_entanglement} by replacing {$\hat{a}_H$ and $\hat{a}_V$ with $\hat{a}_e$ and $\hat{a}_l$, respectively.
\begin{figure}
    \centering
    \includegraphics[width=1.03\linewidth]{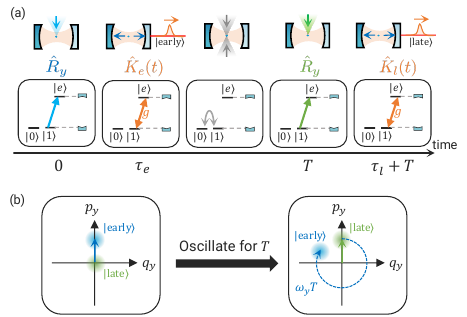}
    \caption{State-dependent emission of time-bin encoded photons.
    (a) Sequence for generating the entanglement between the spin state and the time-bin photon. 
    (b) State-dependent excitation and free oscillation in the harmonic trap result in detrimental entanglement between the motional state and the time-bin photon, if the time-bin separation $T$ is not exactly at an integer multiple of $2\pi/\omega$, as illustrated with simplified representations of Wigner functions in the laboratory frame.
    }
    \label{fig:time_bin_schematic}
\end{figure}
Here $\hat{a}_{e(l)}$ is the annihilation operator of the early (late) time-bin photon.
BSM can be performed by using a single beamsplitter and a pair of single-photon detectors, as shown in Fig.~\ref{fig:infid_timebin}(a), positioned at the central node between two remote atom-cavity systems, enabling projections such as $\hat{\mathcal{P}}(\tau_e,\tau_l) = \hat{a}^+_e(\tau_e)\hat{a}^+_l(\tau_l)$ that establish entanglement between the two atoms~\cite{Barrett2005}.

The atom-photon entanglement generation follows an excitation laser driving $\ket{1}_s \leftrightarrow \ket{e}_s$ transition, whose recoil effect is described by an operator acting on the motional state $\hat{R}_y$.
\begin{figure}[t]
    \centering
    \includegraphics[width=0.85\linewidth]{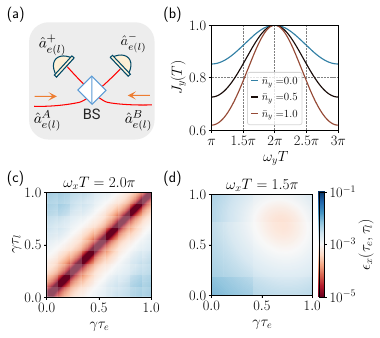}
    \caption{Motion-induced infidelity of the entangled atom pairs generated with the time-bin-based HEG protocol.
    (a) Bell-state measurement setup for the time-bin photons.
    (b) Excitation-laser-induced recoil effect $J_y(T)$ given by Eq.~\eqref{eq:J_y_for_displacement} with $\eta_y = 0.2$.
    Deviation from the condition that $\omega_y T$ is an integer multiple of $2\pi$ results in $J_y < 1$, with larger infidelity for the higher initial temperature.
    (c) Emission-induced infidelity $\epsilon_x(\tau_e,\tau_l)$ for $\omega_x T = 2 \pi$ and (d) $\omega_x T=1.5\pi$.
    The parameters are $(g, \kappa_\text{ex}, \kappa_\text{in}, \omega_{x}) = (10, 100, 1, 0.1)\gamma$, $\eta_{x} = 0.2$ and $\bar{n}_{x}=1$. 
    Configuring $\omega_x T$ and $\omega_y T$ to be integer multiples of $2\pi$ ensures high fidelity.
    }
    \label{fig:infid_timebin}
\end{figure}
\begin{figure*}[t]
	\centering\includegraphics[width=0.99\textwidth]{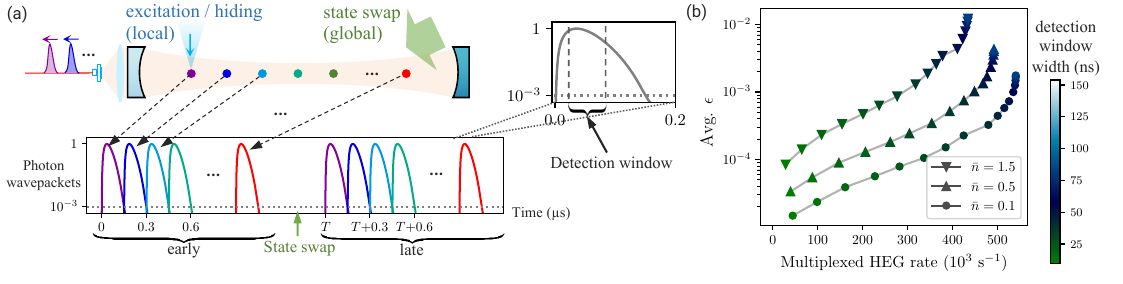}
	\caption{An example implementation of high-rate, high-fidelity operations mediated by time-bin encoded photons.
		(a) Time-mltiplexed cavity-assisted HEG \cite{Huie2021, Li2024, Sunami2025}. 
		Multiple atoms of the same species are coupled to the cavity mode for time-efficient operation in the presence of atom transport and initialization time costs of 100 $\mu$s \cite{Li2024}.
		Addressing laser beams are used to selectively excite one atom at a time, generating photons to be interfered with and detected at the measurement apparatus (Figs.~\ref{fig:time_bin_schematic}(a) and \ref{fig:infid_timebin}(a)) while decoupling the other atoms from the cavity mode to avoid crosstalk.
		The bottom panel illustrates the pulse train from the multiple-atom cavity system (normalized photon wavepackets), where colors indicate the source atom labels. 
		We obtain the overall time cost by considering the case where pulses are separated in time such that the photon wavepacket normalized by their peak values (see plot) decays to $10^{-3}$ before the next pulse is generated, to ensure negligible overlap. 
		(b) The rate-fidelity tradeoff by varying detection windows. 
		For a given window width and initial motional state (thermal state with mean phonon number $\bar{n}$), we obtain the window-integrated photon detection probability and the weighted- average infidelity inside the detection window. 
		The time-multiplexed HEG rate is obtained from the pulse length and heralding probability $p_\mathrm{HEG}=p_e^2/2$, where $p_e$ is the photon generation probability inside the detection window.
	}
	\label{fig:mux}
\end{figure*}
The dynamics of the atom-cavity system with no atomic and cavity decays is described by the non-Hermitian Hamiltonian
\begin{equation}
    \begin{aligned}
	\hat{\mathcal{H}}_\text{t}(t) =& \hbar g \cos[\eta_x (\hat{b}_x e^{-i\omega_x t}+\hat{b}_x^\dagger e^{i\omega_x t})](\hat{c} \ketbra{e}[_s]{1} + \hat{c}^\dagger \ketbra{1}[_s]{e})  \\
	& - i\hbar (\gamma\ketbra{e}[_s]{e} + \kappa \hat{c}^\dagger \hat{c}).
    \end{aligned}
\end{equation}
For the early time-bin photon, the emission-induced recoil effect is given by (see Appendix~\ref{ap:derivation_of_kick_operators} for detailed derivation)
\begin{equation}
    \hat{K}_e(t) = \sqrt{2\kappa_\text{ex}}\tensor[_s]{\bra{1}_c\bra{0}}{}\hat{c}\,\hat{O}_{\mathcal{H}_\text{t}}(t;0)\ket{e}_s\ket{0}_c.
\end{equation}
After the state swap\footnote{
We consider the state swap pulses are applied in a recoil-free manner either by two-photon Raman coupling \cite{Jenkins2022} or with composite pulses \cite{Lis2023, Zhang2024}; thus we safely ignore their effect on the motional states of the atoms. 
The incorporation of the recoil effect from state swaps into the model is straightforward. }, photon emission for the late time bin starts with another excitation laser at time $T$, where the recoil-induced effect is given by
\begin{equation}
    \hat{K}_l(t) = \sqrt{2\kappa_\text{ex}} \tensor[_s]{\bra{1}_c\bra{0}}{}\hat{c}\,\hat{O}_{\mathcal{H}_\text{t}}(t+T;T)\ket{e}_s\ket{0}_c.
\end{equation}
In the frame rotating at the trap frequencies, the recoil effect induced by the excitation at $t=T$ is given by $\hat{U}_{\text{f},y}^\dagger(T) \hat{R}_y \hat{U}_{\text{f},y}(T)$, where $\hat{U}_{\text{f},\mu}(t) = e^{-i\omega_\mu \hat{b}_\mu^\dagger \hat{b}_\mu t}$ represents the free oscillatory evolution in the trap along axis $\mu$; 
the overall kick operators are then
\begin{equation}
    \begin{aligned}
        \hat{\mathcal{K}}_0(t) =& \hat{K}_e(t)\hat{R}_y, \\
        \hat{\mathcal{K}}_1(t) =& \hat{K}_l(t)\hat{U}_{\text{f},y}^\dagger(T) \hat{R}_y \hat{U}_{\text{f},y}(T).
    \end{aligned}
\end{equation}
Assuming the initial motional state along each axis is independent, i.e., $\hat{\rho}_m = \hat{\rho}_{m,x}\otimes \hat{\rho}_{m,y} \otimes \hat{\rho}_{m,z}$, we find that the motion-induced infidelity is given by $\epsilon(\tau_e,\tau_l) = [1- J_y(T)J_x(\tau_e,\tau_l)]/2$, where 
\begin{equation} \label{eq:Jy_timebin}
    J_y(T) = \ab|\Tr\ab[\hat{U}_{\text{f},y}^\dagger(T) \hat{R}_y^\dagger \hat{U}_{\text{f},y}(T)\hat{R}_y\hat{\rho}_{m,y}]|^2,
\end{equation}
and the detection-time-dependent term $J_x(\tau_e,\tau_l)$ has the same form as Eq.~\eqref{eq:J_x} by respectively replacing $\hat{K}_{+}(\tau_{H})$ and $\hat{K}_{-}(\tau_{V})$ with $\hat{K}_{e}(\tau_{e})$ and $\hat{K}_{l}(\tau_{l})$.
Thus, the fidelity depends on not only detection times but also the time separation $T$.

Tuning the trap frequency $\omega_y$ and time separation $T$ such that $\omega_y T$ is an integer multiple of $2\pi$ ensures that $J_y(T)$ reaches unity for any $\hat{R}_y$ and $\hat{\rho}_{m,y}$, because $\hat{U}_{\text{f},y}(t)$ becomes an identity operator.
With the displacement-operator expression for the excitation-induced recoil kick $\hat{R}_y = e^{i\eta_y (\hat{b}_y^\dagger + \hat{b}_y)}$~\cite{Poyatos1996} and the initial state $\hat{\rho}_{m,y}$ being a thermal state with the mean phonon number $\bar{n}_y$, an analytical expression can be obtained as follows (see Appendix~\ref{ap:derivation of J_y}):
\begin{equation} \label{eq:J_y_for_displacement}
    J_y(T) = e^{-2\eta_y^2 (2\bar{n}_y + 1)(1- \cos\omega_y T)},
\end{equation}
which we plot in Fig.~\ref{fig:infid_timebin}(b).
As evident from the expression and the plot, setting $\omega_y T=2\pi$ ensures $J_y(T) = 1$ independent of the mean phonon number. 

As a result of the tuning $\omega_y T\in 2\pi \mathbb{N}$, the motion-induced infidelity now depends only on the motional degree of freedom along $x$ axis: $\epsilon(\tau_e,\tau_l) = \epsilon_x(\tau_e,\tau_l) \coloneqq [1-J_x(\tau_e,\tau_l)]/2$.
Further, it is beneficial to tune $\omega_x T$ to be an integer multiple of $2\pi$, such that $\hat{K}_e(t) = \hat{K}_l(t)$ because $\hat{\mathcal{H}}_\text{t}(t+T)=\hat{\mathcal{H}}_\text{t}(t)$.
This also improves the fidelity, as shown in Figs.~\ref{fig:infid_timebin}(c) and (d).
A similar feature was observed in the time-bin-photon-mediated HEG experiment with free-space photon collection rather than a cavity \cite{Saha2024}, where the trap period condition had to be satisfied for all three axes.
This is in contrast to the cavity-based approach discussed here, where the cavity restricts the direction of the emission-induced kick only to the $x$ direction.
This simplifies the requirement for the protocol design and is practical for atom trapping techniques with intrinsic symmetry of the potentials, such as optical tweezers with cylindrical symmetry where the axes with the same trapping frequencies (e.g.,~$x$ and $y$ directions with $\omega_x = \omega_y$) can be aligned to the cavity mode and the excitation axis.

\section{Time-multiplexed operation}
\label{sec:mux}
Here, we apply the presented model for designing quantum networking operations.
In particular, we consider time-multiplexed HEG operation \cite{Huie2021, Sunami2025, Li2024} mediated by time-bin encoded photons, as illustrated in Fig.~\ref{fig:mux}(a).
The time multiplexing proceeds by preparing a large array of atoms or ions in the cavity mode followed by sequential HEG trials for each atom while applying AC Stark beams to shift the other atoms away from resonance \cite{Li2024, Sunami2025}; this mitigates the large temporal cost associated with atom transport and initialization, resulting in efficient usage of the optical channels for high-rate quantum networking.
Compared to the polarization-based protocols \cite{Li2024}, achieving both high-rate and high-fidelity operation is nontrivial with time-bin-encoded protocols due to the additional requirement for synchronization with atomic motion, as discussed in Sec.~\ref{subsec:time-bin}.
Our model allows the design and evaluation of such an operation.
In Fig.~\ref{fig:mux}, we illustrate our protocol design and resulting performance; we place the photon emissions from one of the internal states in a single batch (early), with minimal time cost to shift the second batch (late) for the other internal states starting at $T$ at an integer multiple of trap periods of both $x$ and $y$ directions which we assume to be the same ($\omega_x = \omega_y \eqqcolon \omega$), as appropriate for optical tweezer traps.
With this, the excitation-induced spin-motion coupling can be minimized (Fig.~\ref{fig:infid_timebin}).
The state swap pulses between the two time bins can be applied simultaneously for all qubits in the cavity, with negligible time cost \cite{Huie2021}.

Figure~\ref{fig:mux}(b) is the predicted trade-off relation between motion-induced infidelity and time-multiplexed HEG rates (see Ref.~\cite{Sunami2025} and Appendix \ref{app:time-mux}). 
The detection-time filtering technique enables HEG rates ranging from $3\times10^5~\mathrm{s}^{-1}$ ($\bar{n}>1$) to $5\times10^5~\mathrm{s}^{-1}$ ($\bar{n}\sim0.1$), while maintaining the average infidelity significantly below $10^{-2}$.
Along with the increased infidelity, the larger $\bar{n}$ results in the slower photon emission process due to the spread of population into multiple motional levels, which effectively decreases the atom-cavity coupling strength \cite{Canteri2024}.
The chosen parameters for this simulation are $(g,\kappa_\text{ex}, \kappa_\text{in} ,\gamma, \omega)/2\pi= (5.0, 7.5, 0.20, 0.20, 0.10)$~\unit{\MHz} with $\eta_x = 0.20$, 
a typical regime for experimentally relevant neutral-atom cavity-QED platforms, such as nanofiber cavity-QED systems, and the atom number in the cavity is 200.
This result, in conjunction with the analysis for other types of infidelities in cavity-assisted quantum interconnect such as in Ref.~\cite{Li2024}, enables the design of high-fidelity quantum networking operations.

\section{conclusion and outlook}
\label{sec:conclusion}
In conclusion, we have presented a comprehensive and tractable model to analyze the recoil-induced infidelity in the cavity-assisted remote entanglement generation protocols, demonstrating that high-fidelity and high-rate operation are possible by careful protocol design including the detection-time filtering.
In particular, cavity-assisted protocols analyzed in this work have allowed the simplification of the motional-mode treatment thanks to the well-defined emission mode and the associated recoil directions.
Our work simplifies and streamlines the analysis of high-fidelity quantum networking operations with scalable trapped-atom qubit systems.

Our model further supports the on-the-fly estimation of 
motion-induced infidelity in each Bell pair, from the photon detection-time information, in a similar spirit as Ref.~\cite{Li2024}.
This can be fed into protocols implemented at the logical level of quantum error-correcting codes, such as decoding in the logical-level entanglement-distillation protocols \cite{Sunami2025, Pattison2024}.
Further system-level optimization for multiprocessor fault-tolerant quantum computing or quantum networking nodes is possible by incorporating the identified rate-fidelity tradeoff.
We leave the physical-to-logical interface design and optimization for future work.

\textit{Note}: In complementary work, Apol\'in and Nadlinger apply the kick-operator model to free-space photon emission without a cavity and propose an alternative method to mitigate the recoil-induced infidelity~\cite{Apolin2025}.

\section*{Declaration of Competing Interest}
S. Kikura, R. Inoue, H. Yamasaki and S. Sunami are employees, and A. Goban is a co-founder and a shareholder of Nanofiber Quantum Technologies, Inc.

\begin{acknowledgements}
We thank V.~Vuleti\'c, H.~Konishi, and K.~N.~Komagata for providing feedback on the manuscript.
\end{acknowledgements}

\appendix

\setcounter{table}{0}
\setcounter{figure}{0}
\renewcommand{\thetable}{A\arabic{table}}
\renewcommand{\thefigure}{A\arabic{figure}}
\renewcommand{\theHtable}{Supplement.\thetable}
\renewcommand{\theHfigure}{Supplement.\thefigure}

\section{General Expression of the Recoil-Induced Infidelity} \label{ap:derivation_of_infid}
In this section, we outline the derivation of our general model for HEG protocols with two-photon interference that incorporates the motional degree of freedom of atoms.
The initial state of the atom, possessing both spin and motional states, is given by
\begin{equation}
    \hat{\rho}_\text{init} = \ketbra{\text{init}}[_s]{\text{init}} \otimes \hat{\rho}_m,
\end{equation}
where initial spin and motional states are $\ketbra{\text{init}}[_s]{\text{init}}$ and $\hat{\rho}_m$, respectively.
Subsequently, the spin-photon entanglement is generated by a state-dependent photon emission.
We model this process by an evolution operator $\hat{\mathcal{G}}$ of spin-motion-photon states after the photon generation process,
\begin{equation} \label{eq:ap_definition_of_mathcal_g}
    \begin{aligned}
        \hat{\mathcal{G}} =& \frac{\ab(\ket{0}_s (\hat{a}_{0}[\hat{\mathcal{K}}_0(t)])^\dagger + \ket{1}_s (\hat{a}_1[\hat{\mathcal{K}}_1(t)])^\dagger )\tensor[_s]{\bra{\text{init}}}{}}{\sqrt{2}},
    \end{aligned}
\end{equation}
where 
\begin{equation}
    (\hat{a}_i[\hat{\mathcal{K}}_i(t)])^\dagger = \int \dd{t} \hat{\mathcal{K}}_i(t) (\hat{a}_i(t))^\dagger,
\end{equation}
using the non-unitary kick operator $\hat{\mathcal{K}}_i(t)$ acting on a motional state. 
Then, the resulting state of the atom and the propagating mode of the emitted photon is given by
\begin{equation} \label{eq:ap_defition_rho}
    \begin{aligned}
        \hat{\rho} = ~ & \hat{\mathcal{G}} (\hat{\rho}_\text{init} \otimes \ketbra{\text{vac}}{\text{vac}}) \hat{\mathcal{G}}^\dagger + \hat{\rho}_\text{fail},
    \end{aligned}
\end{equation}
where $\hat{\rho}_\text{fail}$ represents the unnormalized state corresponding to the event of failed photon emission with the failure probability $\Tr[\hat{\rho}_\text{fail}]$.

Next, we consider the case where both Alice and Bob generate spin-photon entanglement at their respective nodes, and transmit the emitted photons to the BSM apparatus for remote entanglement generation.
Upon obtaining a detection pattern that confirms the success of the BSM, the entanglement between Alice and Bob's spins is generated.
As a representative example of the detection pattern illustrated in Figs. \ref{fig:pol_heg} and \ref{fig:infid_timebin}(a) and discussed in the main text, we define the corresponding positive operator-valued measure (POVM) $\hat{\mathcal{D}}(\tau_0,\tau_1)$ by
\begin{equation}
    \begin{aligned}
        \hat{\mathcal{D}}(\tau_0,\tau_1) =~ & \hat{\mathcal{P}}^\dagger(\tau_0,\tau_1)\hat{\mathcal{P}}(\tau_0,\tau_1), \\
        \hat{\mathcal{P}}(\tau_0,\tau_1) =~ & \hat{a}^{+}_0(\tau_0)\hat{a}^{+}_1(\tau_1),
    \end{aligned}
\end{equation}
where $\hat{a}^{\pm}_i(t) = (\hat{a}^{A}_i(t) \pm \hat{a}^{B}_i(t))/\sqrt{2}$ and  the detection time $\tau_i$.
The superscripts $q\in\{A,B\}$ of the states and operators represent Alice and Bob, respectively. 
From that POVM, we obtain the projected spin-motion states as 
\begin{equation}
    \begin{aligned}
        \hat{\rho}_{sm}(\tau_0,\tau_1) =& \frac{\Tr_p[\hat{\mathcal{D}}(\tau_0,\tau_1) \hat{\rho}^A\otimes \hat{\rho}^B]}{\Tr[\hat{\mathcal{D}}(\tau_0,\tau_1) \hat{\rho}^A\otimes \hat{\rho}^B]},
    \end{aligned}
\end{equation}
where $\Tr_p[\cdot]$ represents the partial trace of propagating modes.
To further simplify the expression, we introduce an operator $\hat{E}(\tau_0, \tau_1)$ for the remote entanglement generation by
\begin{equation} \label{eq:remote_entanglement_op}
    \begin{aligned}
        &\hat{E}(\tau_0,\tau_1) \\
        =& \frac{\ket{01}_s \hat{\mathcal{K}}_0^A(\tau_0)\hat{\mathcal{K}}_1^B(\tau_1) + \ket{10}_s \hat{\mathcal{K}}_1^A(\tau_1)\hat{\mathcal{K}}_0^B(\tau_0)}{4} {}^A_s{\bra{\text{init}}}{}_s^B\bra{\text{init}},
    \end{aligned}
\end{equation}
which projects the spin states onto the Hilbert space consisting of $\{\ket{01}_s,\ket{10}_s\}$ and satisfies the following relation:
\begin{equation}
    \hat{E}(\tau_0,\tau_1)\ket{\text{vac}} = \hat{\mathcal{P}}(\tau_0,\tau_1)\hat{\mathcal{G}}^A\hat{\mathcal{G}}^B\ket{\text{vac}}.
\end{equation}
Thus, the projected spin-motion state is simplified by
\begin{equation}
    \begin{aligned}
        \hat{\rho}_{sm}(\tau_0,\tau_1) 
        =& \frac{\hat{E}(\tau_0,\tau_1) (\hat{\rho}_\text{init}^A\otimes \hat{\rho}_\text{init}^B) \hat{E}^\dagger(\tau_0,\tau_1)}{P_\text{D}(\tau_0,\tau_1)}.
    \end{aligned}
\end{equation}
Here, $P_\text{D}(\tau_0,\tau_1)$ is the probability density of the detection given by
\begin{equation} \label{eq:general_P_D}
    \begin{aligned}
        P_\text{D}(\tau_0,\tau_1) =& \Tr[\hat{E}(\tau_0,\tau_1) (\hat{\rho}_\text{init}^A\otimes \hat{\rho}_\text{init}^B) \hat{E}^\dagger(\tau_0,\tau_1)] \\
        =& \frac{C_{00}^A C_{11}^B + C_{11}^A C_{00}^B}{16},
    \end{aligned}
\end{equation}
where $C_{ij}^q = \Tr[\hat{\mathcal{K}}_i^q(\tau_i)\hat{\rho}_m^q (\hat{\mathcal{K}}_j^q(\tau_j))^\dagger]$.

To evaluate the fidelity of remote spin-spin entanglement, we further trace out the motional state and obtain
\begin{equation}
    \hat{\rho}_s(\tau_0,\tau_1) = \frac{1}{C_{00}^A C_{11}^B + C_{11}^A C_{00}^B} \mqty{C_{00}^A C_{11}^B & C_{01}^A C_{10}^B \\ C_{10}^A C_{01}^B & C_{11}^A C_{00}^B},
\end{equation}
where the basis of the matrix is $\{\ket{01}_s,\ket{10}_s\}$.
Thus, the infidelity from $\ket{\Psi^+}$ is given by
\begin{equation} \label{eq:general_epsilon}
    \epsilon(\tau_0,\tau_1) = \frac{1}{2}-\frac{\Re[C_{01}^A C_{10}^B]}{C_{00}^A C_{11}^B + C_{11}^A C_{00}^B}.
\end{equation}
Assuming that the systems of Alice and Bob are identical, i.e., $\hat{\mathcal{K}}_{0(1)}^{A} = \hat{\mathcal{K}}_{0(1)}^{B} \eqqcolon \hat{\mathcal{K}}_{0(1)}$ and $\hat{\rho}_m^A = \hat{\rho}_m^B \eqqcolon \hat{\rho}_m$, we find
\begin{equation}
    \begin{aligned}
        \hat{\rho}_s(\tau_0,\tau_1) =& (1-\epsilon(\tau_0,\tau_1))\ketbra*{\Psi^{+}}{\Psi^{+}} \\
         &+ \epsilon(\tau_0,\tau_1) \ketbra*{\Psi^-}{\Psi^-}, \\
         \epsilon(\tau_0,\tau_1) =& \frac{1-J(\tau_0,\tau_1)}{2},
    \end{aligned}
\end{equation}
where 
\begin{equation} 
    J(\tau_0,\tau_1) = \frac{\Tr[\hat{\mathcal{K}}_{0}(\tau_0)\hat{\rho}_m\hat{\mathcal{K}}^\dagger_{1}(\tau_1)]\Tr[\hat{\mathcal{K}}_{1}(\tau_1)\hat{\rho}_m\hat{\mathcal{K}}^\dagger_{0}(\tau_0)]}{\Tr[\hat{\mathcal{K}}_{0}(\tau_0)\hat{\rho}_m\hat{\mathcal{K}}^\dagger_{0}(\tau_0)]\Tr[\hat{\mathcal{K}}_{1}(\tau_1)\hat{\rho}_m\hat{\mathcal{K}}^\dagger_{1}(\tau_1)]}.
\end{equation}
Note that while the other detection pattern $\{\hat{a}_0^-(\tau_0), \hat{a}_1^{-}(\tau_1)\}$ projects the initial spin state onto $\ket{\Psi^{+}}$, the other patterns: $\{\hat{a}_0^+(\tau_0), \hat{a}_1^{-}(\tau_1)\}$ and $\{\hat{a}_0^-(\tau_0), \hat{a}_1^{+}(\tau_1)\}$ project the spins onto $\ket{\Psi^-}$ with the same infidelity $\epsilon(\tau_0,\tau_1)$ and probability density $P_\text{D}(\tau_0,\tau_1)$; Thus the success probability density of the protocol is $4P_\text{D}(\tau_0,\tau_1)$.

This model has a wide range of applicability, covering not only cavity-enhanced photon generation but also free-space emission.
Given the kick operator $\hat{\mathcal{K}}_i^{q}$, derived by a direct time evolution of the photon generation process, we can straightforwardly evaluate the success probability and the recoil-induced infidelity of a HEG protocol, by using Eqs.~\eqref{eq:general_P_D} and \eqref{eq:general_epsilon}.

\section{Derivation of Kick Operators}
\label{ap:derivation_of_kick_operators}
In this section, we describe the detailed derivation of the kick operators.
We first consider the polarization-encoding case, for which the dynamics during the photon emission is governed by the Hamiltonian~\eqref{eq:pol_Hamiltonian}.
To identify kick operators, we assume the initial state to be $\ket{e}_s\ket{0}_c\ket{\phi}_m$, where $\ket{0}_c$ is a vacuum state of two cavity modes and $\ket{\phi}_m$ is an arbitrary motional state.
We first consider the dynamics of the atom-cavity system under the condition of no atomic and cavity field decays.
Here, we regard the propagating modes as the environment, to which the cavity mode leaks out with rate $\kappa_\mathrm{ex}$.
The dynamics is described by the non-Hermitian Schr\"{o}dinger equation:
\begin{equation}
    i\hbar \odv{\ket{\psi(t)}}{t} = \hat{\mathcal{H}}_\text{p}(t)\ket{\psi(t)},
\end{equation}
where $\hat{\mathcal{H}}_\text{p}(t) \coloneqq \hat{H}_\text{p}(t) - i\hbar[\gamma \ketbra{e}[_s]{e}+ \kappa(\hat{c}_+^\dagger\hat{c}_+ + \hat{c}_-^\dagger\hat{c}_-)]$~\cite{Law1997,Plenio1998,Goto2019}.
We formally solve this equation by
\begin{equation}
    \ket{\psi(t)} = \hat{O}_{\mathcal{H}_\text{p}}(t;0) \ket{e}_s\ket{0}_c\ket{\phi}_m,
\end{equation}
where 
\begin{equation}
    \hat{O}_{\mathcal{H}_\text{p}}(t;t_0) = \mathcal{T}\ab[\exp\ab(-\frac{i}{\hbar}\int_{t_0}^{t} \dd{t^\prime} \hat{\mathcal{H}}_\text{p}(t^\prime))].
\end{equation}
This intra-cavity state relates to the propagating photonic state as follows~\cite{Goto2019,Krutyanskiy2023,Kikura2024}:
\begin{equation}
    \ket{\zeta} = \int_0^\infty \dd{t} \sqrt{2\kappa_\text{ex}} (\hat{a}_H^\dagger(t) \hat{c}_{+} + \hat{a}_V^\dagger(t) \hat{c}_{-}) \ket{\psi(t)}\ket{\text{vac}},
\end{equation}
where $\ket{\zeta}$ is the state consisting of the spin, cavity mode, motional mode, and propagating mode.
After the photon emission the cavity state remains in $\ket{0}_c$, and thus the spin-motion-photon state is given by ${}_c\braket{0}{\zeta}$.
Setting the ansatz
\begin{equation}
    \begin{aligned} 
    {}_c\braket{0}{\zeta} = \frac{1}{\sqrt{2}} \Big( & \ket{0}_s \int \dd{t} \hat{K}_+(t) \hat{a}_{H}^\dagger(t) \\
    &+ \ket{1}_s \int \dd{t} \hat{K}_-(t) \hat{a}_{V}^\dagger(t) \Big) \ket{\phi}_m \ket{\text{vac}},
    \end{aligned}
\end{equation}
we find
\begin{equation}
    \begin{aligned}
        \hat{K}_+(t)\ket{0}_s\ket{\phi}_m =& 2\sqrt{\kappa_\text{ex}}\,_{c}\bra{0}\hat{c}_+ \hat{O}_{\mathcal{H}_\text{p}}(t;0)\ket{e}_s \ket{0}_c\ket{\phi}_m, \\
        \hat{K}_-(t)\ket{1}_s\ket{\phi}_m =& 2\sqrt{\kappa_\text{ex}} \,_{c}\bra{0}\hat{c}_- \hat{O}_{\mathcal{H}_\text{p}}(t;0)\ket{e}_s \ket{0}_c\ket{\phi}_m.
    \end{aligned}
\end{equation}
This result shows that for any initial motional state $\hat{\rho}_m$ the emission-induced kick operators are given by
\begin{equation}
    \begin{aligned}
        \hat{K}_{+}(t) =& 2\sqrt{\kappa_\text{ex}} \tensor[_s]{\bra{0}_c\bra{0}}{} \hat{c}_{+}\hat{O}_{\mathcal{H}_\text{p}}(t;0)\ket{e}_s\ket{0}_c, \\
        \hat{K}_{-}(t) =& 2\sqrt{\kappa_\text{ex}} \tensor[_s]{\bra{1}_c\bra{0}}{} \hat{c}_{-}\hat{O}_{\mathcal{H}_\text{p}}(t;0)\ket{e}_s\ket{0}_c.
    \end{aligned}
\end{equation}

We also consider the time-bin encoding case, where the Hamiltonian is given by 
\begin{equation}
    \begin{aligned}
        \hat{H}_\text{t}(t) = &\hbar g \cos[\eta_x (\hat{b}_x e^{-i\omega_x t}+\hat{b}_x^\dagger e^{i\omega_x t})] \\
        & \times (\hat{c} \ketbra{e}[_s]{1} + \hat{c}^\dagger \ketbra{1}[_s]{e}).
    \end{aligned}
\end{equation}
With the initial state $(\ket{0}_s+\ket{1}_s)\ket{0}_c\ket{\phi}_m/\sqrt{2}$, we first apply the state-dependent excitation whose evolution operator is given by
\begin{equation}
    \widehat{\text{Ex}} = \ketbra{0}[_s]{0}\otimes \ketbra{0}[_c]{0} \otimes \hat{I}_{\mathcal{M}}+ \ketbra{e}[_s]{1}\otimes \ketbra{0}[_c]{0} \otimes \hat{R}_y,
\end{equation}
where $\hat{I}_{\mathcal{M}}$ is the identity operator of the motional degree of freedom.
Subsequently, the spin-motion-cavity state under the dynamics with no atomic and cavity decays is given by
\begin{equation}
    \begin{aligned}
        &\hat{O}_{\mathcal{H}_\text{t}}(t;0) \widehat{\text{Ex}} (\ket{0}_s+\ket{1}_s)\ket{0}_c\ket{\phi}_m/\sqrt{2} \\
        =&\frac{\ket{0}_s\ket{0}_c\ket{\phi}_m + \hat{O}_{\mathcal{H}_\text{t}}(t;0) \hat{R}_y\ket{e}_s\ket{0}_c \ket{\phi}_m}{\sqrt{2}},
    \end{aligned}
\end{equation}
where we have defined $\hat{\mathcal{H}}_\text{t}(t) = \hat{H}_\text{t}(t) - i\hbar (\gamma\ketbra{e}[_s]{e} + \kappa \hat{c}^\dagger\hat{c})$.
Thus, the total state including the propagating mode after the first photon generation is given by
\begin{equation*}
    \ab(\ket{0}_s\ket{0}_c + \ket{1}_s\ket{0}_c \int\dd{t} \hat{K}_e(t) \hat{R}_y \hat{a}_e^\dagger(t))\ket{\phi}_m\ket{\text{vac}}/\sqrt{2},
\end{equation*}
where the kick operator is given by
\begin{equation}
    \hat{K}_e(t) = \sqrt{2\kappa_\text{ex}}\tensor[_s]{\bra{1}_c\bra{0}}{}\hat{c}\,\hat{O}_{\mathcal{H}_\text{t}}(t;0)\ket{e}_s\ket{0}_c.
\end{equation}
After we apply the state swap or the Pauli X gate, and wait to set the time-separation to be $T$, we apply the state-dependent excitation again.
In the frame rotating at trap frequencies, the second excitation is described by
\begin{equation*}
    \begin{aligned}
        \widehat{\text{Ex}}^{(l)} =& \hat{U}_\text{f}^\dagger(T)\widehat{\text{Ex}}\hat{U}_\text{f}(T) \\
        =& \ketbra{0}[_s]{0}\otimes \ketbra{0}[_c]{0} \otimes \hat{I}_{\mathcal{M}}+ \ketbra{e}[_s]{1}\otimes \ketbra{0}[_c]{0} \otimes \hat{R}_y^{(l)},
    \end{aligned}
\end{equation*}
where $\hat{U}_\text{f}(t) = \Pi_{\mu = {x,y,z}} \hat{U}_{\text{f},\mu}(t)$ and $\hat{R}_y^{(l)} = \hat{U}_{\text{f},y}^\dagger(T)\hat{R}_y\hat{U}_{\text{f},y}(T)$.
Thus, the final state after the late photon generation is given by
\begin{equation}
    \begin{aligned}
        \frac{1}{\sqrt{2}} \Big(& \ket{0}_s\ket{0}_c\int\dd{t} \hat{K}_e(t) \hat{R}_y \hat{a}_e^\dagger(t)\ket{\phi}_m\ket{\text{vac}} \\
        & + \ket{1}_s\ket{0}_c\int\dd{t} \hat{K}_l(t) \hat{R}_y^{(l)}\hat{a}_l^\dagger(t) \ket{\phi}_m\ket{\text{vac}}\Big),
    \end{aligned}
\end{equation}
where 
\begin{equation}
    \hat{K}_l(t) = \sqrt{2\kappa_\text{ex}}\tensor[_s]{\bra{1}_c\bra{0}}{}\hat{c}\,\hat{O}_{\mathcal{H}_\text{t}}(t+T;T)\ket{e}_s\ket{0}_c.
\end{equation}
We identify the overall kick operators as follows:
\begin{equation}
    \begin{aligned}
        \hat{\mathcal{K}}_0(t) =& \hat{K}_e(t) \hat{R}_y, \\
        \hat{\mathcal{K}}_1(t) =& \hat{K}_l(t) \hat{R}_y^{(l)},
    \end{aligned}
\end{equation}
for any initial motional state $\hat{\rho}_m$.

\section{Numerical Simulation of the Time-Ordered Operator} \label{ap:numerical_calculation_method_of_K(t)}
We show an efficient calculation method for time-ordered operators $\hat{O}_{\mathcal{H}_\text{p}}(t;t_0)$ for polarization encoding and $\hat{O}_{\mathcal{H}_\text{t}}(t;t_0)$ for time-bin encoding.
We first discuss $\hat{O}_{\mathcal{H}_\text{p}}(t;t_0)$.
A brute-force calculation method is to discretize the time with the sufficiently small time step $\Delta t_j = t_{j+1}-t_j$ and approximate the operator as
\begin{equation} \label{eq.ap_naive_calc_time_order_operator}
    \hat{O}_{\mathcal{H}_\text{p}}(t_{N+1};t_0) \approx e^{-i\hat{\mathcal{H}}_p(t_N)\Delta t_N/\hbar} \cdots e^{-i\hat{\mathcal{H}}_p(t_0)\Delta t_0/\hbar}.
\end{equation}
In contrast, one can derive a rigorous and computationally efficient expression of $\hat{O}_{\mathcal{H}_\text{p}}(t;t_0)$ by utilizing the fact that the Hamiltonian $\hat{H}_\text{p}(t)$ on the rotating frame of trap frequencies  is derived from the time-independent one~\eqref{eq:H_w/o_rotaing_frame_pol} as described in the following:
We consider the unnormalized state vector $\ket{\psi(t)} =  \hat{O}_{\mathcal{H}_\text{p}}(t;t_0) \ket{\psi(t_0)}$ with the arbitrary pure state $\ket{\psi(t_0)}$.
The temporal dynamics of the state vector is described by solving the Shr\"{o}dinger equation given by
\begin{equation} \label{eq:app_shrodinger_eq_rotating}
    i\hbar \odv{\ket{\psi(t)}}{t} = \hat{\mathcal{H}}_\text{p}(t)\ket{\psi(t)}.
\end{equation}
Here, we consider the Hamiltonian given by
\begin{equation}
    \begin{aligned}
        \hat{\bar{H}}_\text{p} =& \sum_{\mu=x,y,z}\hbar \omega_\mu \hat{b}_\mu^\dagger\hat{b}_\mu  \\
         & +\hbar g\cos[\eta_x(\hat{b}_x +\hat{b}_x^\dagger)] (\hat{c}_+ \ketbra{e}[_s]{0} + \hat{c}_- \ketbra{e}[_s]{1} + \text{H.c.}),
    \end{aligned}
\end{equation}
and $\hat{H}_\text{p}(t)$ is derived from $\hat{\bar{H}}_\text{p}$ by way of
\begin{equation} \label{eq:relation_between_lab_rotating_frame}
    \hat{H}_\text{p}(t) = \hat{U}_\text{f}^\dagger(t)\hat{\bar{H}}_\text{p} \hat{U}_\text{f}(t) + i\hbar \ab(\odv{\hat{U}_\text{f}^\dagger(t)}{t}) \hat{U}_\text{f}(t), 
\end{equation}
where $\hat{U}_\text{f}(t) = e^{-i \sum_{\mu=x,y,z}\omega_\mu \hat{b}_\mu^\dagger\hat{b}_\mu t}$ represents the free oscillatory evolution.
From Eqs.~\eqref{eq:app_shrodinger_eq_rotating} and \eqref{eq:relation_between_lab_rotating_frame}, we find the Schr\"{o}dinger equation for the transformed state $\ket{\bar\psi(t)} = \hat{U}_\text{f}(t)\ket{\psi(t)}$ given by
\begin{equation} 
    i\hbar \odv{\ket{\bar{\psi}(t)}}{t} = \hat{\bar{\mathcal{H}}}_\text{p} \ket*{\bar{\psi}(t)},
\end{equation}
where $\hat{\bar{\mathcal{H}}}_\text{p} = \hat{\bar{H}}_\text{p}- i\hbar[\gamma \ketbra{e}[_s]{e}+ \kappa (\hat{c}_+^\dagger\hat{c}_+ + \hat{c}_-^\dagger\hat{c}_-)]$.
This equation gives $\ket{\bar{\psi}(t)} = e^{-i\hat{\bar{\mathcal{H}}}_\text{p}(t-t_0)/\hbar}\ket{\bar{\psi}(t_0)}$, and thus we find
\begin{equation}
    \ket{\psi(t)} = \hat{U}_\text{f}^\dagger(t) e^{-i\hat{\bar{\mathcal{H}}}_\text{p}(t-t_0)/\hbar} \hat{U}_\text{f}(t_0) \ket{\psi(t_0)}.
\end{equation}
By comparing it with $\ket{\psi(t)} =  \hat{O}_{\mathcal{H}_\text{p}}(t;t_0) \ket{\psi(t_0)}$, we obtain
\begin{equation}  \label{eq:O_pol}
    \hat{O}_{\mathcal{H}_\text{p}}(t;t_0) = \hat{U}_\text{f}^\dagger(t) e^{-i\hat{\bar{\mathcal{H}}}_\text{p}(t-t_0)/\hbar} \hat{U}_\text{f}(t_0).
\end{equation}

For $\hat{O}_{\mathcal{H}_\text{t}}(t;t_0)$, we can use the same recipe and obtain
\begin{equation} \label{eq:O_time_bin}
    \hat{O}_{\mathcal{H}_\text{t}}(t;t_0) = \hat{U}_\text{f}^\dagger(t) e^{-i\hat{\bar{\mathcal{H}}}_\text{t}(t-t_0)/\hbar} \hat{U}_\text{f}(t_0),
\end{equation}
where
\begin{equation*}
	\begin{aligned}
		\hat{\bar{\mathcal{H}}}_\text{t}
        = & \sum_{\mu=x,y,z}\hbar \omega_\mu \hat{b}_\mu^\dagger\hat{b}_\mu \\
        & + \hbar g \cos[\eta_x (\hat{b}_x+\hat{b}_x^\dagger)] (\hat{c} \ketbra{e}[_s]{1} + \hat{c}^\dagger \ketbra{1}[_s]{e})  \\
	& - i\hbar (\gamma\ketbra{e}[_s]{e} + \kappa \hat{c}^\dagger \hat{c}).
	\end{aligned}
\end{equation*}
The results shown in the main text are obtained by numerically calculating Eqs.~\eqref{eq:O_pol} and \eqref{eq:O_time_bin} with QuTiP~\cite{Johansson2013}.

\begin{figure}[t]
    \centering
    \includegraphics[width=\linewidth]{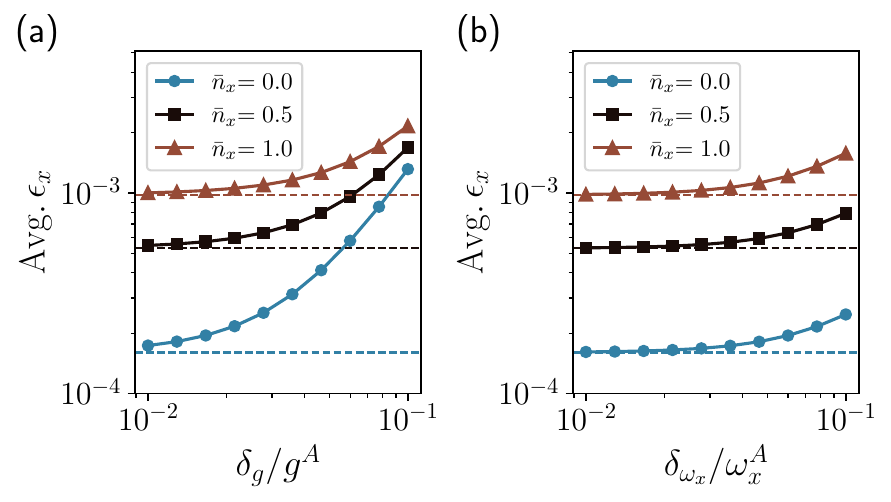}
    \caption{Motion-induced infidelity of the time-bin-encoding HEG protocol in the presence of difference in system parameters between the two modules, while the following parameters are the same: 
    $(\kappa_{\text{ex}}, \kappa_\text{in}, \omega_y) = (100, 1, 0.1)\gamma$ and $\eta_x = 0.2$.
    We also consider the case $T = 2\pi/\omega_y$ for both cavities. 
    The parameters $g$ and $\omega_x$ are varied between the two systems as $g^B = g^A + \delta_g$ and $\omega_x^B = \omega_x^A + \delta_{\omega_x}$ with $(g^A,\omega_x^A) = (10, 0.1)\gamma$.
    The initial motional states of two modules are characterized by the same motional quanta $\bar{n}_x$.
    (a) Average infidelity as a function of $\delta_g/g^A$ for $\delta_{\omega_x} = 0$. 
    The dotted lines illustrate the case for $\delta_g = 0$.
    A few \% level deviation of $g$, combined with near-ground state cooling $\bar{n}_x<0.5$, suppresses $\epsilon_x$.
    (b) Average infidelity as a function of $\delta_{\omega_x}/\omega_x^A$ for $\delta_g = 0$.
    The dotted lines represent the infidelity for $\delta_{\omega_x} = 0$.
    Here, the dependence on $\delta_{\omega_x}$ is moderate, tolerating $\sim 10$ \% deviation at $\bar{n}<0.5$.
    }
\label{fig:infidelity_from_inhomogeneity}
\end{figure}
\section{Motional-State Dependent Atom-Photon Coupling }
\label{ap:motion_dep_g}
Assuming a trap frequency is sufficiently larger than the bandwidth of a cavity, the transition probability between different motional eigenstates $\ket{n_x}_{m,x}$ of $\hat{b}_x^\dagger\hat{b}_x$ through photon generation is negligible.
In this motional-sideband resolved regime, the effective atom-photon coupling rate can be defined for each motional levels.
In the frame rotating at the trap frequency $\omega_x$, the position-dependent coupling strength is given as
\begin{equation}
    \begin{aligned}
        &g\cos[\eta_x(\hat{b}_x e^{-i\omega_x t} + \hat{b}_x^\dagger e^{i\omega_x t})] \\
        =& g \ab[1-\frac{\eta_x^2}{2}(\hat{b}_x\hat{b}_x^\dagger + \hat{b}_x^2e^{-2i\omega_x t} + \text{H.c.})] + \mathcal{O}(\eta_x^4) \\
        \simeq& g \ab[1-\frac{\eta_x^2}{2}(2\hat{b}^\dagger_x \hat{b}_x + 1)] + \mathcal{O}(\eta_x^4),
    \end{aligned}
\end{equation}
where we have used $[\hat{b}_x, \hat{b}_x^\dagger] = 1$ and neglected the higher-frequency components $\hat{b}_x^2e^{-2i\omega_x t}$ and $\hat{b}_x^{^\dagger 2}e^{2i\omega_x t}$, called the rotating-wave approximation.
Thus, effective coupling strength for the motional quanta $n_x$ is given by
\begin{eqnarray}
    g_{n_x} \simeq g[1-\eta_x^2(n_x + 1/2)],
\end{eqnarray}
up to the third order of $\eta_x$.
This means that the finite temperature of the trapped atom reduces the indistinguishability of the photon, leading to increased infidelity. 

\section{Infidelity from the System Parameter Difference Between the Two Cavities}
\label{ap:infidelity_from_parameter_difference}
In the main article, we have assumed that Alice and Bob have identical systems, which is desired for high-fidelity HEG.
In a realistic situation, however, there should be fractional differences between system parameters, which increases the infidelity.
Our model can incorporate such inhomogeneity by constructing (different) kick operators of Alice and Bob, as explained in Appendix~\ref{ap:derivation_of_infid}.
As an example, we first consider that the atom-photon coupling strength $g^B$ of Bob is slightly larger than that of Alice: $g^B = g^A + \delta_g$.
This may arise from the nonzero variation of the cavity mode distribution \cite{Li2024} or the difference in the finesse of two cavities.
As shown in Fig.~\ref{fig:infidelity_from_inhomogeneity}(a), this disparity introduces a rapid increase in the infidelity averaged over the entire probability distribution.

The discrepancy of the $x$-axis trap frequencies between the two parties may also arise, resulting in $\omega_x^B = \omega_x^A + \delta_{\omega_x}$.
As shown in Fig.~\ref{fig:infidelity_from_inhomogeneity}(b), as long as the excitation-laser induced error $J_y$ is managed by the appropriate choice of $T$, $\delta_{\omega_x}$ does not result in a significant effect.

\section{Running-Wave Cavities}
\label{ap:running_wave_cavity}
We investigate the scaling of infidelity with the Lamb-Dicke parameter for the standing-wave cavity, as detailed in the main article, and a running-wave cavity.
In the running-wave cavity, atomic transitions $\ket{0}_s \leftrightarrow \ket{e}_s$ and $\ket{1}_s \leftrightarrow \ket{e}_s$ are respectively coupled to the cavity modes $\hat{c}_+$ and $\hat{c}_-$, in a similar manner to the standing-wave cavity in Sec.~\ref{sec:kick_pol}.
First, we consider the case where two running modes copropagate along $x$ axis, e.g., realized by a twisted ring cavity~\cite{Schine2016,Nigyuan2018,Li2024}.
The Hamiltonian is given by
\begin{equation}
    \hat{H}_\text{run}(t) = \hbar g [\hat{\text{d}}_x(t)(\hat{c}_+ \ketbra{e}[_s]{0} + \hat{c}_- \ketbra{e}[_s]{1}) + \text{H.c.} ],
\end{equation}
where $\hat{\text{d}}_x(t) = e^{i\eta_x(\hat{b}_x e^{-i\omega_x t} + \hat{b}_x^\dagger e^{i\omega_x t})}$.
As shown in Fig. \ref{fig:ring_eta_scale}, the infidelity for the running-wave cavity with unidirectional photon emission follows LD parameter scaling  $\epsilon \propto\eta_x^2$, whereas it follows $\epsilon \propto\eta_x^4$ for the standing-wave cavity, regardless of whether the system is in the sideband-resolved or unresolved regime.
This different scaling arises because the running-wave cavity allows the motional coupling with  $\Delta n_x=\pm 1$, while the standing-wave cavity forbids it.
\begin{figure}[t]
    \centering
    \includegraphics[width=\linewidth]{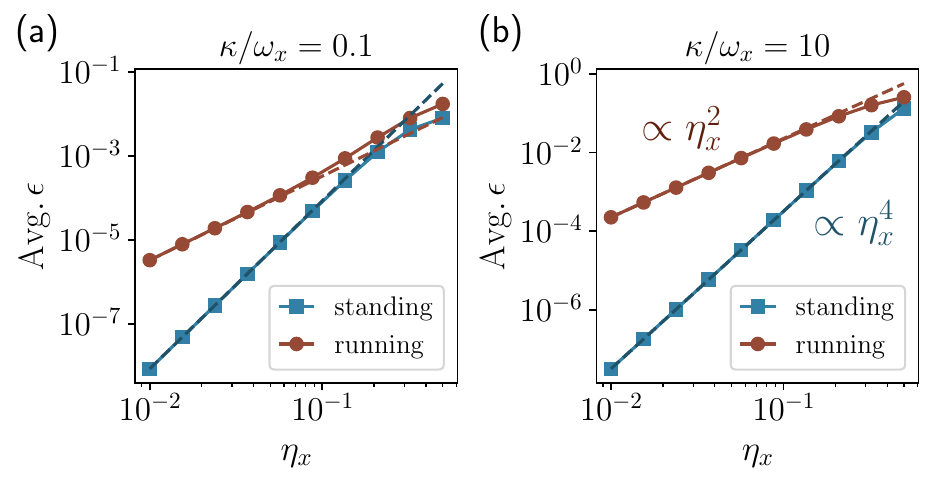}
    \caption{Lamb-Dicke parameter scaling of the average infidelity for the standing- and running-wave cavities at (a) the resolved $(\kappa/\omega_x = 0.1)$ and (b) unresolved $(\kappa/\omega_x = 10)$ regime. 
    The parameters are $(g,\kappa_\text{ex},\kappa_\text{in}) = (10, 100, 1)\gamma$ and $\bar{n}_x=1$.
    The infidelities of standing- and running-wave cavity respectively scale as $\eta_x^4$ and $\eta_x^2$ for small $\eta_x$.}
    \label{fig:ring_eta_scale}
\end{figure}
Note that in the case where two running modes counterpropagate, we find that the fidelity is significantly degraded, because the directions of the photon-recoil kick in motional phase space should be almost opposite for two polarization modes regardless of the detection times.

\section{Fast-Excitation Approximation}
\label{ap:separation_excitation_photon_gen}
In the main text, we have set $\Omega(t) \gg r_c \coloneqq \min(\kappa, g^2/\kappa)$, i.e., the excitation of an atom to $\ket{e}_s$ is much faster than the effective bandwidth of photon generation based on cavity QED~\cite{Utsugi2022}, which has allowed us to treat the fast excitation and the photon-generation dynamics as independent.
Here, we perform a simplified analysis to identify a concrete applicability range of this approximation.
We consider a two-level system $\{\ket{g}_s, \ket{e}_s\}$ that resonantly couples to the cavity mode $\hat{c}$.
The Hamiltonian is 
\begin{equation}
    \begin{aligned}
        \hat{H}(t) =& \sum_{\mu = x,y} \hbar \omega_\mu \hat{b}_\mu^\dagger \hat{b}_\mu + \Omega(t)\ab[\hat{\sigma}_+ e^{i\eta_y (\hat{b}_y^\dagger + \hat{b}_y)} + \text{H.c.}] \\
        & +\hbar g \cos[\eta_x(\hat{b}_x^\dagger + \hat{b}_x)] (\hat{c}^\dagger \hat{\sigma}_- + \hat{c} \hat{\sigma}_+),
    \end{aligned}
\end{equation}
where $\hat{\sigma}_+ = \ketbra{e}[_s]{g}$ and $\hat{\sigma}_- = \ketbra{g}[_s]{e}$.
We consider the dynamics where we apply a $\pi$ pulse with the Rabi frequency $\Omega$ in time $0\leq t \leq t_0 \coloneqq \pi/(2\Omega)$ and numerically solve the master equation:
\begin{equation}
    \odv{\hat{\rho}}{t} =-\frac{i}{\hbar}[\hat{H}(t), \hat{\rho}] + \sum_i \ab(\hat{L}_i \hat{\rho} \hat{L}_i^\dagger - \frac{1}{2}\{\hat{L}^\dagger \hat{L}, \hat{\rho}\}),
\end{equation}  
with the Lindblad operators $\hat{L}_0 = \sqrt{2\kappa}\hat{c}$ and $\hat{L}_1= \sqrt{2\gamma}\hat{\sigma}_-$, to get the state $\hat{\rho}(t_0)$ after the excitation.
We evaluate the overlap between the simulated final state and the ideal state $\hat{\rho}_{\mathrm{id}}(t_0)$ where two dynamics are completely independent, which can be simulated by setting $g=0$.
The overlap is defined as the fidelity
\begin{equation}
    F_\text{overlap} = \ab(\Tr\sqrt{\sqrt{\hat{\rho}(t_0)}\hat{\rho}_{\mathrm{id}}(t_0) \sqrt{\hat{\rho}(t_0)}} )^2.
\end{equation}
\begin{figure}[t]
    \centering
    \includegraphics[width=0.5\linewidth]{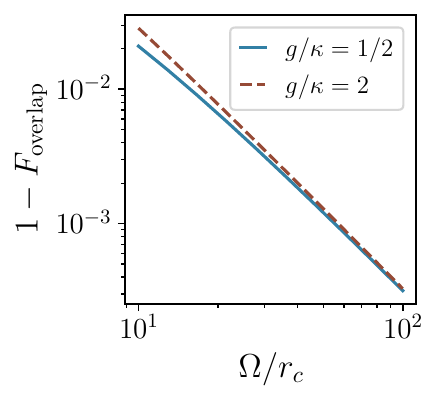}
    \caption{Deviation of the spin-motion state from the 
    fast-excitation limit $\Omega \gg r_c = \min(\kappa, g^2/\kappa)$.
    The parameters are $(g, \omega_x, \omega_y) = (5.0, 0.10, 0.10)\gamma$ and $\eta_x = \eta_y =0.20$.
    We have used $r_c = g^2/\kappa = 2.5\gamma$ for $g/\kappa = 1/2$, and $r_c = \kappa = 2.5\gamma$ for $g/\kappa = 2$.
    }
    \label{fig:infidelity_for_fast-excitation_app}
\end{figure}
As shown in Fig.~\ref{fig:infidelity_for_fast-excitation_app}, $1-F_\text{overlap}$ is sufficiently suppressed for the realistic values of $\Omega$, such as those satisfying $\Omega/r_c$ exceeding several tens.
We note that our model is also applicable for the case of a slower excitation with mixed dynamics along $x$ and $y$ axes;
this can be done by constructing the kick operators from a full-Hamiltonian dynamics including the excitation lasers, e.g., by using Eq.~\eqref{eq.ap_naive_calc_time_order_operator}, and with expression~\eqref{eq:general_epsilon}.
Concrete evaluation of such a situation is outside of the scope of this paper.

\section{Derivation of Eq.~\eqref{eq:J_y_for_displacement}} \label{ap:derivation of J_y}
For a simple derivation, we express the initial thermal state in a coherent-state basis as follows~\cite{Glauber1963}:
\begin{equation}
    \hat{\rho}_{m,y} = \frac{1}{\pi \bar{n}_y} \int \d^2\alpha e^{-|\alpha|^2/\bar{n}_y} \ketbra{\alpha}[_{m,y}]{\alpha}.
\end{equation}
For $\hat{R}_y = e^{i\eta_y(\hat{b}_y^\dagger + \hat{b}_y)}$, we find 
\begin{equation*}
    \begin{aligned}
        \hat{U}_{\text{f},y}^\dagger(T) \hat{R}_y^\dagger \hat{U}_{\text{f},y}(T)\hat{R}_y =& \hat{D}_y(-i\eta_y e^{i\omega_y T})\hat{D}_y(i\eta_y) \\
        =& e^{-i\eta_y^2 \sin\omega_yT}\hat{D}_y[i\eta_y(1-e^{i\omega_y T})],
    \end{aligned}
\end{equation*}
where we have defined the displacement operator $\hat{D}_y(\beta) = e^{\beta\hat{b}_y^\dagger - \beta^\ast \hat{b}_y}$.
From this relation, we obtain
\begin{equation}
    \begin{aligned}
        &J_y(T) \\
        =& \ab|\frac{1}{\pi \bar{n}_y} \int \d^2\alpha e^{-|\alpha|^2/\bar{n}_y}\ _{m,y}\braket*[3]{\alpha}{\hat{D}_y[i\eta_y(1-e^{i\omega_y T})]}{\alpha}_{m,y}|^2 \\
        =& \ab|\frac{e^{-\eta_y^2|1-e^{i\omega_yT}|^2/2}}{\pi \bar{n}_y} \int \d^2\alpha e^{-|\alpha|^2/\bar{n}_y} e^{i\eta_y(1-e^{i\omega_y T})\alpha^\ast - \text{H.c.}}|^2 \\
        =& e^{-2\eta_y^2(2\bar{n}_y + 1)(1-\cos\omega_yT)}.
    \end{aligned}
\end{equation}

\section{Time-Multiplexed Operations}
\label{app:time-mux}
In Fig.~\ref{fig:mux}, we analyze the time-multiplexed protocol for time-bin-photon-mediated HEG as in Refs.~\cite{Huie2021,Sunami2025}, with additional considerations of the requirement to choose the time separation $T$ between the time bins such that $\omega T$ is an integer multiple of $2\pi$ in order to suppress the recoil effects (Fig.~\ref{fig:infid_timebin}).
Here, in contrast to the long-distance quantum communication considered in Ref.~\cite{Huie2021}, we focus on the short-distance entanglement generation for multiprocessor quantum computing as in Refs.~\cite{Li2024,Sunami2025}, where the fidelity is a highly crucial metric; we thus neglect the channel losses and transmission time overhead analyzed in Ref.~\cite{Huie2021}.

In the following, we provide the detail for calculating the realistic HEG rate with $N$ atoms within the cavity, in the presence of time costs of atom transport in and out of the cavity $t_\text{move}$, atom state initialization $t_\text{init}$, and state swap to be applied between two batches $t_\text{swap}$, as we used for Fig.~\ref{fig:mux}(b).
First, following the atom transport and state preparation with $t_\text{move} + t_\text{init}$, sequential photon emission for each atom in $\ket{1}$ is performed with time cost $N t_\text{pulse}$, where $t_\text{pulse}$ is the pulse length as defined in the main text.
After performing the state swap for $t_\text{swap}$ in a recoil-free manner, it is then necessary to wait until the nearest integer multiple of $\omega T/2\pi$ from the first photon emission, to perform the second batch of photon emission lasting for another $N t_\text{pulse}$.
In other words, the total execution time $T$ of the first batch of photon generation operations, between the end of state initialization and the start of the second batch, is given by
\begin{equation} \label{eq:T_rep}
    T = \frac{2\pi}{\omega} \Bigg\lceil \frac{\omega(Nt_\text{pulse} + t_\text{swap})}{2\pi} \Bigg\rceil.
\end{equation}
Sequential photon emission operations, with the same repetition frequency as the first batch, ensure the suppression of the recoil effect for all atoms, as analyzed in the main text. 
Successful HEG trials are labelled, such that atoms successfully ending up in remote Bell states are not involved in the next round of the HEG operations for remaining $N(1-p^2_e/2)$ atoms.
In this round, the appropriate wait time between the first and second batches of photon emissions may be different.

Repeating the above procedure for $M$ rounds, we end up with $N_\text{Bell}^{(M)} = \sum_{m=1}^{M} N_m p_e^2/2$ Bell pairs on average, where $N_m$ is the number of HEG trials in round $m$, which satisfy $N_m = N_{m-1}(1-p_e^2/2)$ and $N_1 = N$.
The associated time cost is $t_M = t_\text{move} +\sum_{m=1}^M (t_\text{init} + T_m + N_m t_\text{pulse})$, where $T_m$ is given by replacing $N$ with $N_m$ in Eq.~\eqref{eq:T_rep}.
For $(N, t_\text{move}, t_\text{init}, t_\text{swap}) = (200, \qty{100}{\us}, \qty{20}{\us}, \qty{1}{\us})$ and the system parameters described in Sec.~\ref{sec:mux}, we numerically evaluate $N_\text{Bell}^{(M)}$ and $t_M$ for $M$ ranging between 1 and 60, and report the maximum value of the rate $ N_\text{Bell}^{(M)}/t_M$ in Fig.~\ref{fig:mux}(b).
The optimal $M$ is typically around 10, which can be reduced significantly by zoned operations within the cavity with increased parallelism, albeit with added experimental complexity \cite{Sunami2025}.

\bibliography{refs}

\providecommand{\noopsort}[1]{}\providecommand{\singleletter}[1]{#1}%
\begin{thebibliography}{54}%
\makeatletter
\providecommand \@ifxundefined [1]{%
 \@ifx{#1\undefined}
}%
\providecommand \@ifnum [1]{%
 \ifnum #1\expandafter \@firstoftwo
 \else \expandafter \@secondoftwo
 \fi
}%
\providecommand \@ifx [1]{%
 \ifx #1\expandafter \@firstoftwo
 \else \expandafter \@secondoftwo
 \fi
}%
\providecommand \natexlab [1]{#1}%
\providecommand \enquote  [1]{``#1''}%
\providecommand \bibnamefont  [1]{#1}%
\providecommand \bibfnamefont [1]{#1}%
\providecommand \citenamefont [1]{#1}%
\providecommand \href@noop [0]{\@secondoftwo}%
\providecommand \href [0]{\begingroup \@sanitize@url \@href}%
\providecommand \@href[1]{\@@startlink{#1}\@@href}%
\providecommand \@@href[1]{\endgroup#1\@@endlink}%
\providecommand \@sanitize@url [0]{\catcode `\\12\catcode `\$12\catcode `\&12\catcode `\#12\catcode `\^12\catcode `\_12\catcode `\%12\relax}%
\providecommand \@@startlink[1]{}%
\providecommand \@@endlink[0]{}%
\providecommand \url  [0]{\begingroup\@sanitize@url \@url }%
\providecommand \@url [1]{\endgroup\@href {#1}{\urlprefix }}%
\providecommand \urlprefix  [0]{URL }%
\providecommand \Eprint [0]{\href }%
\providecommand \doibase [0]{https://doi.org/}%
\providecommand \selectlanguage [0]{\@gobble}%
\providecommand \bibinfo  [0]{\@secondoftwo}%
\providecommand \bibfield  [0]{\@secondoftwo}%
\providecommand \translation [1]{[#1]}%
\providecommand \BibitemOpen [0]{}%
\providecommand \bibitemStop [0]{}%
\providecommand \bibitemNoStop [0]{.\EOS\space}%
\providecommand \EOS [0]{\spacefactor3000\relax}%
\providecommand \BibitemShut  [1]{\csname bibitem#1\endcsname}%
\let\auto@bib@innerbib\@empty
\bibitem [{\citenamefont {Kimble}(2008)}]{Kimble2008}%
  \BibitemOpen
  \bibfield  {author} {\bibinfo {author} {\bibfnamefont {H.~J.}\ \bibnamefont {Kimble}},\ }\bibfield  {title} {\bibinfo {title} {The quantum internet},\ }\href {https://doi.org/10.1038/nature07127} {\bibfield  {journal} {\bibinfo  {journal} {Nature}\ }\textbf {\bibinfo {volume} {453}},\ \bibinfo {pages} {1023} (\bibinfo {year} {2008})}\BibitemShut {NoStop}%
\bibitem [{\citenamefont {Wehner}\ \emph {et~al.}(2018)\citenamefont {Wehner}, \citenamefont {Elkouss},\ and\ \citenamefont {Hanson}}]{Wehner2018}%
  \BibitemOpen
  \bibfield  {author} {\bibinfo {author} {\bibfnamefont {S.}~\bibnamefont {Wehner}}, \bibinfo {author} {\bibfnamefont {D.}~\bibnamefont {Elkouss}},\ and\ \bibinfo {author} {\bibfnamefont {R.}~\bibnamefont {Hanson}},\ }\bibfield  {title} {\bibinfo {title} {Quantum internet: A vision for the road ahead},\ }\href {https://www.science.org/doi/10.1126/science.aam9288} {\bibfield  {journal} {\bibinfo  {journal} {Science}\ }\textbf {\bibinfo {volume} {362}} (\bibinfo {year} {2018})}\BibitemShut {NoStop}%
\bibitem [{\citenamefont {Azuma}\ \emph {et~al.}(2023)\citenamefont {Azuma}, \citenamefont {Economou}, \citenamefont {Elkouss}, \citenamefont {Hilaire}, \citenamefont {Jiang}, \citenamefont {Lo},\ and\ \citenamefont {Tzitrin}}]{Azuma2023}%
  \BibitemOpen
  \bibfield  {author} {\bibinfo {author} {\bibfnamefont {K.}~\bibnamefont {Azuma}}, \bibinfo {author} {\bibfnamefont {S.~E.}\ \bibnamefont {Economou}}, \bibinfo {author} {\bibfnamefont {D.}~\bibnamefont {Elkouss}}, \bibinfo {author} {\bibfnamefont {P.}~\bibnamefont {Hilaire}}, \bibinfo {author} {\bibfnamefont {L.}~\bibnamefont {Jiang}}, \bibinfo {author} {\bibfnamefont {H.-K.}\ \bibnamefont {Lo}},\ and\ \bibinfo {author} {\bibfnamefont {I.}~\bibnamefont {Tzitrin}},\ }\bibfield  {title} {\bibinfo {title} {Quantum repeaters: From quantum networks to the quantum internet},\ }\href {https://doi.org/10.1103/RevModPhys.95.045006} {\bibfield  {journal} {\bibinfo  {journal} {Rev. Mod. Phys.}\ }\textbf {\bibinfo {volume} {95}},\ \bibinfo {pages} {045006} (\bibinfo {year} {2023})}\BibitemShut {NoStop}%
\bibitem [{\citenamefont {Covey}\ \emph {et~al.}(2023)\citenamefont {Covey}, \citenamefont {Weinfurter},\ and\ \citenamefont {Bernien}}]{Covey2023}%
  \BibitemOpen
  \bibfield  {author} {\bibinfo {author} {\bibfnamefont {J.~P.}\ \bibnamefont {Covey}}, \bibinfo {author} {\bibfnamefont {H.}~\bibnamefont {Weinfurter}},\ and\ \bibinfo {author} {\bibfnamefont {H.}~\bibnamefont {Bernien}},\ }\bibfield  {title} {\bibinfo {title} {Quantum networks with neutral atom processing nodes},\ }\href {https://doi.org/10.1038/s41534-023-00759-9} {\bibfield  {journal} {\bibinfo  {journal} {npj Quantum Information 2023 9:1}\ }\textbf {\bibinfo {volume} {9}},\ \bibinfo {pages} {1} (\bibinfo {year} {2023})}\BibitemShut {NoStop}%
\bibitem [{\citenamefont {Young}\ \emph {et~al.}(2022)\citenamefont {Young}, \citenamefont {Safari}, \citenamefont {Huft}, \citenamefont {Zhang}, \citenamefont {Oh}, \citenamefont {Chinnarasu},\ and\ \citenamefont {Saffman}}]{Young2022}%
  \BibitemOpen
  \bibfield  {author} {\bibinfo {author} {\bibfnamefont {C.~B.}\ \bibnamefont {Young}}, \bibinfo {author} {\bibfnamefont {A.}~\bibnamefont {Safari}}, \bibinfo {author} {\bibfnamefont {P.}~\bibnamefont {Huft}}, \bibinfo {author} {\bibfnamefont {J.}~\bibnamefont {Zhang}}, \bibinfo {author} {\bibfnamefont {E.}~\bibnamefont {Oh}}, \bibinfo {author} {\bibfnamefont {R.}~\bibnamefont {Chinnarasu}},\ and\ \bibinfo {author} {\bibfnamefont {M.}~\bibnamefont {Saffman}},\ }\bibfield  {title} {\bibinfo {title} {An architecture for quantum networking of neutral atom processors},\ }\href {https://doi.org/10.1007/S00340-022-07865-0/TABLES/4} {\bibfield  {journal} {\bibinfo  {journal} {Applied Physics B: Lasers and Optics}\ }\textbf {\bibinfo {volume} {128}},\ \bibinfo {pages} {1} (\bibinfo {year} {2022})}\BibitemShut {NoStop}%
\bibitem [{\citenamefont {Monroe}\ \emph {et~al.}(2014)\citenamefont {Monroe}, \citenamefont {Raussendorf}, \citenamefont {Ruthven}, \citenamefont {Brown}, \citenamefont {Maunz}, \citenamefont {Duan},\ and\ \citenamefont {Kim}}]{Monroe2014}%
  \BibitemOpen
  \bibfield  {author} {\bibinfo {author} {\bibfnamefont {C.}~\bibnamefont {Monroe}}, \bibinfo {author} {\bibfnamefont {R.}~\bibnamefont {Raussendorf}}, \bibinfo {author} {\bibfnamefont {A.}~\bibnamefont {Ruthven}}, \bibinfo {author} {\bibfnamefont {K.~R.}\ \bibnamefont {Brown}}, \bibinfo {author} {\bibfnamefont {P.}~\bibnamefont {Maunz}}, \bibinfo {author} {\bibfnamefont {L.-M.}\ \bibnamefont {Duan}},\ and\ \bibinfo {author} {\bibfnamefont {J.}~\bibnamefont {Kim}},\ }\bibfield  {title} {\bibinfo {title} {Large-scale modular quantum-computer architecture with atomic memory and photonic interconnects},\ }\href {https://doi.org/10.1103/PhysRevA.89.022317} {\bibfield  {journal} {\bibinfo  {journal} {Phys. Rev. A}\ }\textbf {\bibinfo {volume} {89}},\ \bibinfo {pages} {022317} (\bibinfo {year} {2014})}\BibitemShut {NoStop}%
\bibitem [{\citenamefont {Sinclair}\ \emph {et~al.}()\citenamefont {Sinclair}, \citenamefont {Ramette}, \citenamefont {Grinkemeyer}, \citenamefont {Bluvstein}, \citenamefont {Lukin},\ and\ \citenamefont {Vuleti^^c4^^87}}]{Sinclair2024}%
  \BibitemOpen
  \bibfield  {author} {\bibinfo {author} {\bibfnamefont {J.}~\bibnamefont {Sinclair}}, \bibinfo {author} {\bibfnamefont {J.}~\bibnamefont {Ramette}}, \bibinfo {author} {\bibfnamefont {B.}~\bibnamefont {Grinkemeyer}}, \bibinfo {author} {\bibfnamefont {D.}~\bibnamefont {Bluvstein}}, \bibinfo {author} {\bibfnamefont {M.}~\bibnamefont {Lukin}},\ and\ \bibinfo {author} {\bibfnamefont {V.}~\bibnamefont {Vuleti^^c4^^87}},\ }\bibfield  {title} {\bibinfo {title} {Fault-tolerant optical interconnects for neutral-atom arrays},\ }\href {https://arxiv.org/abs/2408.08955} {\ }\Eprint {https://arxiv.org/abs/2408.08955} {arXiv:2408.08955 [quant-ph]} \BibitemShut {NoStop}%
\bibitem [{\citenamefont {Sunami}\ \emph {et~al.}(2025)\citenamefont {Sunami}, \citenamefont {Tamiya}, \citenamefont {Inoue}, \citenamefont {Yamasaki},\ and\ \citenamefont {Goban}}]{Sunami2025}%
  \BibitemOpen
  \bibfield  {author} {\bibinfo {author} {\bibfnamefont {S.}~\bibnamefont {Sunami}}, \bibinfo {author} {\bibfnamefont {S.}~\bibnamefont {Tamiya}}, \bibinfo {author} {\bibfnamefont {R.}~\bibnamefont {Inoue}}, \bibinfo {author} {\bibfnamefont {H.}~\bibnamefont {Yamasaki}},\ and\ \bibinfo {author} {\bibfnamefont {A.}~\bibnamefont {Goban}},\ }\bibfield  {title} {\bibinfo {title} {Scalable networking of neutral-atom qubits: Nanofiber-based approach for multiprocessor fault-tolerant quantum computers},\ }\href {https://doi.org/10.1103/PRXQuantum.6.010101} {\bibfield  {journal} {\bibinfo  {journal} {PRX Quantum}\ }\textbf {\bibinfo {volume} {6}},\ \bibinfo {pages} {010101} (\bibinfo {year} {2025})}\BibitemShut {NoStop}%
\bibitem [{\citenamefont {Bennett}\ \emph {et~al.}(1999)\citenamefont {Bennett}, \citenamefont {Shor}, \citenamefont {Smolin},\ and\ \citenamefont {Thapliyal}}]{Bennett1999}%
  \BibitemOpen
  \bibfield  {author} {\bibinfo {author} {\bibfnamefont {C.~H.}\ \bibnamefont {Bennett}}, \bibinfo {author} {\bibfnamefont {P.~W.}\ \bibnamefont {Shor}}, \bibinfo {author} {\bibfnamefont {J.~A.}\ \bibnamefont {Smolin}},\ and\ \bibinfo {author} {\bibfnamefont {A.~V.}\ \bibnamefont {Thapliyal}},\ }\bibfield  {title} {\bibinfo {title} {Entanglement-assisted classical capacity of noisy quantum channels},\ }\href {https://doi.org/10.1103/PhysRevLett.83.3081} {\bibfield  {journal} {\bibinfo  {journal} {Phys. Rev. Lett.}\ }\textbf {\bibinfo {volume} {83}},\ \bibinfo {pages} {3081} (\bibinfo {year} {1999})}\BibitemShut {NoStop}%
\bibitem [{\citenamefont {Panayi}\ \emph {et~al.}(2014)\citenamefont {Panayi}, \citenamefont {Razavi}, \citenamefont {Ma},\ and\ \citenamefont {L^^c3^^bctkenhaus}}]{Panayi2014}%
  \BibitemOpen
  \bibfield  {author} {\bibinfo {author} {\bibfnamefont {C.}~\bibnamefont {Panayi}}, \bibinfo {author} {\bibfnamefont {M.}~\bibnamefont {Razavi}}, \bibinfo {author} {\bibfnamefont {X.}~\bibnamefont {Ma}},\ and\ \bibinfo {author} {\bibfnamefont {N.}~\bibnamefont {L^^c3^^bctkenhaus}},\ }\bibfield  {title} {\bibinfo {title} {Memory-assisted measurement-device-independent quantum key distribution},\ }\href {https://doi.org/10.1088/1367-2630/16/4/043005} {\bibfield  {journal} {\bibinfo  {journal} {New Journal of Physics}\ }\textbf {\bibinfo {volume} {16}},\ \bibinfo {pages} {043005} (\bibinfo {year} {2014})}\BibitemShut {NoStop}%
\bibitem [{\citenamefont {Gottesman}\ \emph {et~al.}(2012)\citenamefont {Gottesman}, \citenamefont {Jennewein},\ and\ \citenamefont {Croke}}]{Gottesman2012}%
  \BibitemOpen
  \bibfield  {author} {\bibinfo {author} {\bibfnamefont {D.}~\bibnamefont {Gottesman}}, \bibinfo {author} {\bibfnamefont {T.}~\bibnamefont {Jennewein}},\ and\ \bibinfo {author} {\bibfnamefont {S.}~\bibnamefont {Croke}},\ }\bibfield  {title} {\bibinfo {title} {Longer-baseline telescopes using quantum repeaters},\ }\href {https://doi.org/10.1103/PhysRevLett.109.070503} {\bibfield  {journal} {\bibinfo  {journal} {Phys. Rev. Lett.}\ }\textbf {\bibinfo {volume} {109}},\ \bibinfo {pages} {070503} (\bibinfo {year} {2012})}\BibitemShut {NoStop}%
\bibitem [{\citenamefont {Khabiboulline}\ \emph {et~al.}(2019)\citenamefont {Khabiboulline}, \citenamefont {Borregaard}, \citenamefont {De~Greve},\ and\ \citenamefont {Lukin}}]{Khabiboulline2019}%
  \BibitemOpen
  \bibfield  {author} {\bibinfo {author} {\bibfnamefont {E.~T.}\ \bibnamefont {Khabiboulline}}, \bibinfo {author} {\bibfnamefont {J.}~\bibnamefont {Borregaard}}, \bibinfo {author} {\bibfnamefont {K.}~\bibnamefont {De~Greve}},\ and\ \bibinfo {author} {\bibfnamefont {M.~D.}\ \bibnamefont {Lukin}},\ }\bibfield  {title} {\bibinfo {title} {Optical interferometry with quantum networks},\ }\href {https://doi.org/10.1103/PhysRevLett.123.070504} {\bibfield  {journal} {\bibinfo  {journal} {Phys. Rev. Lett.}\ }\textbf {\bibinfo {volume} {123}},\ \bibinfo {pages} {070504} (\bibinfo {year} {2019})}\BibitemShut {NoStop}%
\bibitem [{\citenamefont {Fitzsimons}(2017)}]{Fitzsimons2017}%
  \BibitemOpen
  \bibfield  {author} {\bibinfo {author} {\bibfnamefont {J.~F.}\ \bibnamefont {Fitzsimons}},\ }\bibfield  {title} {\bibinfo {title} {Private quantum computation: an introduction to blind quantum computing and related protocols},\ }\href {https://doi.org/10.1038/S41534-017-0025-3} {\bibfield  {journal} {\bibinfo  {journal} {npj Quantum Information}\ }\textbf {\bibinfo {volume} {3}},\ \bibinfo {pages} {23} (\bibinfo {year} {2017})}\BibitemShut {NoStop}%
\bibitem [{\citenamefont {Duan}\ and\ \citenamefont {Kimble}(2003)}]{Duan2003}%
  \BibitemOpen
  \bibfield  {author} {\bibinfo {author} {\bibfnamefont {L.-M.}\ \bibnamefont {Duan}}\ and\ \bibinfo {author} {\bibfnamefont {H.~J.}\ \bibnamefont {Kimble}},\ }\bibfield  {title} {\bibinfo {title} {Efficient engineering of multiatom entanglement through single-photon detections},\ }\href {https://doi.org/10.1103/PhysRevLett.90.253601} {\bibfield  {journal} {\bibinfo  {journal} {Phys. Rev. Lett.}\ }\textbf {\bibinfo {volume} {90}},\ \bibinfo {pages} {253601} (\bibinfo {year} {2003})}\BibitemShut {NoStop}%
\bibitem [{\citenamefont {Barrett}\ and\ \citenamefont {Kok}(2005)}]{Barrett2005}%
  \BibitemOpen
  \bibfield  {author} {\bibinfo {author} {\bibfnamefont {S.~D.}\ \bibnamefont {Barrett}}\ and\ \bibinfo {author} {\bibfnamefont {P.}~\bibnamefont {Kok}},\ }\bibfield  {title} {\bibinfo {title} {Efficient high-fidelity quantum computation using matter qubits and linear optics},\ }\href {https://doi.org/10.1103/PhysRevA.71.060310} {\bibfield  {journal} {\bibinfo  {journal} {Phys. Rev. A}\ }\textbf {\bibinfo {volume} {71}},\ \bibinfo {pages} {060310} (\bibinfo {year} {2005})}\BibitemShut {NoStop}%
\bibitem [{\citenamefont {Beukers}\ \emph {et~al.}(2024)\citenamefont {Beukers}, \citenamefont {Pasini}, \citenamefont {Choi}, \citenamefont {Englund}, \citenamefont {Hanson},\ and\ \citenamefont {Borregaard}}]{Beukers2024}%
  \BibitemOpen
  \bibfield  {author} {\bibinfo {author} {\bibfnamefont {H.~K.}\ \bibnamefont {Beukers}}, \bibinfo {author} {\bibfnamefont {M.}~\bibnamefont {Pasini}}, \bibinfo {author} {\bibfnamefont {H.}~\bibnamefont {Choi}}, \bibinfo {author} {\bibfnamefont {D.}~\bibnamefont {Englund}}, \bibinfo {author} {\bibfnamefont {R.}~\bibnamefont {Hanson}},\ and\ \bibinfo {author} {\bibfnamefont {J.}~\bibnamefont {Borregaard}},\ }\bibfield  {title} {\bibinfo {title} {Remote-entanglement protocols for stationary qubits with photonic interfaces},\ }\href {https://doi.org/10.1103/PRXQuantum.5.010202} {\bibfield  {journal} {\bibinfo  {journal} {PRX Quantum}\ }\textbf {\bibinfo {volume} {5}},\ \bibinfo {pages} {010202} (\bibinfo {year} {2024})}\BibitemShut {NoStop}%
\bibitem [{\citenamefont {Ritter}\ \emph {et~al.}(2012)\citenamefont {Ritter}, \citenamefont {N^^c3^^b6lleke}, \citenamefont {Hahn}, \citenamefont {Reiserer}, \citenamefont {Neuzner}, \citenamefont {Uphoff}, \citenamefont {M^^c3^^bccke}, \citenamefont {Figueroa}, \citenamefont {Bochmann},\ and\ \citenamefont {Rempe}}]{Ritter2012}%
  \BibitemOpen
  \bibfield  {author} {\bibinfo {author} {\bibfnamefont {S.}~\bibnamefont {Ritter}}, \bibinfo {author} {\bibfnamefont {C.}~\bibnamefont {N^^c3^^b6lleke}}, \bibinfo {author} {\bibfnamefont {C.}~\bibnamefont {Hahn}}, \bibinfo {author} {\bibfnamefont {A.}~\bibnamefont {Reiserer}}, \bibinfo {author} {\bibfnamefont {A.}~\bibnamefont {Neuzner}}, \bibinfo {author} {\bibfnamefont {M.}~\bibnamefont {Uphoff}}, \bibinfo {author} {\bibfnamefont {M.}~\bibnamefont {M^^c3^^bccke}}, \bibinfo {author} {\bibfnamefont {E.}~\bibnamefont {Figueroa}}, \bibinfo {author} {\bibfnamefont {J.}~\bibnamefont {Bochmann}},\ and\ \bibinfo {author} {\bibfnamefont {G.}~\bibnamefont {Rempe}},\ }\bibfield  {title} {\bibinfo {title} {An elementary quantum network of single atoms in optical cavities},\ }\href {https://doi.org/10.1038/nature11023} {\bibfield  {journal} {\bibinfo  {journal} {Nature}\ }\textbf {\bibinfo {volume} {484}},\ \bibinfo {pages} {195} (\bibinfo {year} {2012})}\BibitemShut {NoStop}%
\bibitem [{\citenamefont {Main}\ \emph {et~al.}(2025)\citenamefont {Main}, \citenamefont {Drmota}, \citenamefont {Nadlinger}, \citenamefont {Ainley}, \citenamefont {Agrawal}, \citenamefont {Nichol}, \citenamefont {Srinivas}, \citenamefont {Araneda},\ and\ \citenamefont {Lucas}}]{Main2024}%
  \BibitemOpen
  \bibfield  {author} {\bibinfo {author} {\bibfnamefont {D.}~\bibnamefont {Main}}, \bibinfo {author} {\bibfnamefont {P.}~\bibnamefont {Drmota}}, \bibinfo {author} {\bibfnamefont {D.}~\bibnamefont {Nadlinger}}, \bibinfo {author} {\bibfnamefont {E.}~\bibnamefont {Ainley}}, \bibinfo {author} {\bibfnamefont {A.}~\bibnamefont {Agrawal}}, \bibinfo {author} {\bibfnamefont {B.}~\bibnamefont {Nichol}}, \bibinfo {author} {\bibfnamefont {R.}~\bibnamefont {Srinivas}}, \bibinfo {author} {\bibfnamefont {G.}~\bibnamefont {Araneda}},\ and\ \bibinfo {author} {\bibfnamefont {D.}~\bibnamefont {Lucas}},\ }\bibfield  {title} {\bibinfo {title} {Distributed quantum computing across an optical network link},\ }\href {https://doi.org/10.1038/s41586-024-08404-x} {\bibfield  {journal} {\bibinfo  {journal} {Nature}\ }\textbf {\bibinfo {volume} {638}},\ \bibinfo {pages} {383} (\bibinfo {year} {2025})}\BibitemShut {NoStop}%
\bibitem [{\citenamefont {Saha}\ \emph {et~al.}()\citenamefont {Saha}, \citenamefont {Shalaev}, \citenamefont {O'Reilly}, \citenamefont {Goetting}, \citenamefont {Toh}, \citenamefont {Kalakuntla}, \citenamefont {Yu},\ and\ \citenamefont {Monroe}}]{Saha2024}%
  \BibitemOpen
  \bibfield  {author} {\bibinfo {author} {\bibfnamefont {S.}~\bibnamefont {Saha}}, \bibinfo {author} {\bibfnamefont {M.}~\bibnamefont {Shalaev}}, \bibinfo {author} {\bibfnamefont {J.}~\bibnamefont {O'Reilly}}, \bibinfo {author} {\bibfnamefont {I.}~\bibnamefont {Goetting}}, \bibinfo {author} {\bibfnamefont {G.}~\bibnamefont {Toh}}, \bibinfo {author} {\bibfnamefont {A.}~\bibnamefont {Kalakuntla}}, \bibinfo {author} {\bibfnamefont {Y.}~\bibnamefont {Yu}},\ and\ \bibinfo {author} {\bibfnamefont {C.}~\bibnamefont {Monroe}},\ }\bibfield  {title} {\bibinfo {title} {High-fidelity remote entanglement of trapped atoms mediated by time-bin photons},\ }\href@noop {} {\ }\Eprint {https://arxiv.org/abs/2406.01761} {arXiv:2406.01761 [quant-ph]} \BibitemShut {NoStop}%
\bibitem [{\citenamefont {O'Reilly}\ \emph {et~al.}(2024)\citenamefont {O'Reilly}, \citenamefont {Toh}, \citenamefont {Goetting}, \citenamefont {Saha}, \citenamefont {Shalaev}, \citenamefont {Carter}, \citenamefont {Risinger}, \citenamefont {Kalakuntla}, \citenamefont {Li}, \citenamefont {Verma},\ and\ \citenamefont {Monroe}}]{Oreilly2024}%
  \BibitemOpen
  \bibfield  {author} {\bibinfo {author} {\bibfnamefont {J.}~\bibnamefont {O'Reilly}}, \bibinfo {author} {\bibfnamefont {G.}~\bibnamefont {Toh}}, \bibinfo {author} {\bibfnamefont {I.}~\bibnamefont {Goetting}}, \bibinfo {author} {\bibfnamefont {S.}~\bibnamefont {Saha}}, \bibinfo {author} {\bibfnamefont {M.}~\bibnamefont {Shalaev}}, \bibinfo {author} {\bibfnamefont {A.~L.}\ \bibnamefont {Carter}}, \bibinfo {author} {\bibfnamefont {A.}~\bibnamefont {Risinger}}, \bibinfo {author} {\bibfnamefont {A.}~\bibnamefont {Kalakuntla}}, \bibinfo {author} {\bibfnamefont {T.}~\bibnamefont {Li}}, \bibinfo {author} {\bibfnamefont {A.}~\bibnamefont {Verma}},\ and\ \bibinfo {author} {\bibfnamefont {C.}~\bibnamefont {Monroe}},\ }\bibfield  {title} {\bibinfo {title} {Fast photon-mediated entanglement of continuously cooled trapped ions for quantum networking},\ }\href {https://doi.org/10.1103/PhysRevLett.133.090802} {\bibfield  {journal} {\bibinfo  {journal} {Phys. Rev. Lett.}\ }\textbf {\bibinfo {volume} {133}},\ \bibinfo {pages}
  {090802} (\bibinfo {year} {2024})}\BibitemShut {NoStop}%
\bibitem [{\citenamefont {Jenkins}\ \emph {et~al.}(2022)\citenamefont {Jenkins}, \citenamefont {Lis}, \citenamefont {Senoo}, \citenamefont {McGrew},\ and\ \citenamefont {Kaufman}}]{Jenkins2022}%
  \BibitemOpen
  \bibfield  {author} {\bibinfo {author} {\bibfnamefont {A.}~\bibnamefont {Jenkins}}, \bibinfo {author} {\bibfnamefont {J.~W.}\ \bibnamefont {Lis}}, \bibinfo {author} {\bibfnamefont {A.}~\bibnamefont {Senoo}}, \bibinfo {author} {\bibfnamefont {W.~F.}\ \bibnamefont {McGrew}},\ and\ \bibinfo {author} {\bibfnamefont {A.~M.}\ \bibnamefont {Kaufman}},\ }\bibfield  {title} {\bibinfo {title} {Ytterbium nuclear-spin qubits in an optical tweezer array},\ }\href {https://doi.org/10.1103/PhysRevX.12.021027} {\bibfield  {journal} {\bibinfo  {journal} {Phys. Rev. X}\ }\textbf {\bibinfo {volume} {12}},\ \bibinfo {pages} {021027} (\bibinfo {year} {2022})}\BibitemShut {NoStop}%
\bibitem [{\citenamefont {Bluvstein}\ \emph {et~al.}(2022)\citenamefont {Bluvstein}, \citenamefont {Levine}, \citenamefont {Semeghini}, \citenamefont {Wang}, \citenamefont {Ebadi}, \citenamefont {Kalinowski}, \citenamefont {Keesling}, \citenamefont {Maskara}, \citenamefont {Pichler}, \citenamefont {Greiner}, \citenamefont {Vuleti^^c4^^87},\ and\ \citenamefont {Lukin}}]{Bluvstein2022}%
  \BibitemOpen
  \bibfield  {author} {\bibinfo {author} {\bibfnamefont {D.}~\bibnamefont {Bluvstein}}, \bibinfo {author} {\bibfnamefont {H.}~\bibnamefont {Levine}}, \bibinfo {author} {\bibfnamefont {G.}~\bibnamefont {Semeghini}}, \bibinfo {author} {\bibfnamefont {T.~T.}\ \bibnamefont {Wang}}, \bibinfo {author} {\bibfnamefont {S.}~\bibnamefont {Ebadi}}, \bibinfo {author} {\bibfnamefont {M.}~\bibnamefont {Kalinowski}}, \bibinfo {author} {\bibfnamefont {A.}~\bibnamefont {Keesling}}, \bibinfo {author} {\bibfnamefont {N.}~\bibnamefont {Maskara}}, \bibinfo {author} {\bibfnamefont {H.}~\bibnamefont {Pichler}}, \bibinfo {author} {\bibfnamefont {M.}~\bibnamefont {Greiner}}, \bibinfo {author} {\bibfnamefont {V.}~\bibnamefont {Vuleti^^c4^^87}},\ and\ \bibinfo {author} {\bibfnamefont {M.~D.}\ \bibnamefont {Lukin}},\ }\bibfield  {title} {\bibinfo {title} {A quantum processor based on coherent transport of entangled atom arrays},\ }\href {https://doi.org/10.1038/s41586-022-04592-6} {\bibfield  {journal} {\bibinfo  {journal} {Nature}\
  }\textbf {\bibinfo {volume} {604}},\ \bibinfo {pages} {451} (\bibinfo {year} {2022})}\BibitemShut {NoStop}%
\bibitem [{\citenamefont {Evered}\ \emph {et~al.}(2023)\citenamefont {Evered}, \citenamefont {Bluvstein}, \citenamefont {Kalinowski}, \citenamefont {Ebadi}, \citenamefont {Manovitz}, \citenamefont {Zhou}, \citenamefont {Li}, \citenamefont {Geim}, \citenamefont {Wang}, \citenamefont {Maskara}, \citenamefont {Levine}, \citenamefont {Semeghini}, \citenamefont {Greiner}, \citenamefont {Vuleti^^c4^^87},\ and\ \citenamefont {Lukin}}]{Evered2023}%
  \BibitemOpen
  \bibfield  {author} {\bibinfo {author} {\bibfnamefont {S.~J.}\ \bibnamefont {Evered}}, \bibinfo {author} {\bibfnamefont {D.}~\bibnamefont {Bluvstein}}, \bibinfo {author} {\bibfnamefont {M.}~\bibnamefont {Kalinowski}}, \bibinfo {author} {\bibfnamefont {S.}~\bibnamefont {Ebadi}}, \bibinfo {author} {\bibfnamefont {T.}~\bibnamefont {Manovitz}}, \bibinfo {author} {\bibfnamefont {H.}~\bibnamefont {Zhou}}, \bibinfo {author} {\bibfnamefont {S.~H.}\ \bibnamefont {Li}}, \bibinfo {author} {\bibfnamefont {A.~A.}\ \bibnamefont {Geim}}, \bibinfo {author} {\bibfnamefont {T.~T.}\ \bibnamefont {Wang}}, \bibinfo {author} {\bibfnamefont {N.}~\bibnamefont {Maskara}}, \bibinfo {author} {\bibfnamefont {H.}~\bibnamefont {Levine}}, \bibinfo {author} {\bibfnamefont {G.}~\bibnamefont {Semeghini}}, \bibinfo {author} {\bibfnamefont {M.}~\bibnamefont {Greiner}}, \bibinfo {author} {\bibfnamefont {V.}~\bibnamefont {Vuleti^^c4^^87}},\ and\ \bibinfo {author} {\bibfnamefont {M.~D.}\ \bibnamefont {Lukin}},\ }\bibfield  {title} {\bibinfo
  {title} {High-fidelity parallel entangling gates on a neutral-atom quantum computer},\ }\href {https://doi.org/10.1038/s41586-023-06481-y} {\bibfield  {journal} {\bibinfo  {journal} {Nature}\ }\textbf {\bibinfo {volume} {622}},\ \bibinfo {pages} {268} (\bibinfo {year} {2023})}\BibitemShut {NoStop}%
\bibitem [{\citenamefont {Peper}\ \emph {et~al.}(2025)\citenamefont {Peper}, \citenamefont {Li}, \citenamefont {Knapp}, \citenamefont {Bileska}, \citenamefont {Ma}, \citenamefont {Liu}, \citenamefont {Peng}, \citenamefont {Zhang}, \citenamefont {Horvath}, \citenamefont {Burgers},\ and\ \citenamefont {Thompson}}]{Peper2025}%
  \BibitemOpen
  \bibfield  {author} {\bibinfo {author} {\bibfnamefont {M.}~\bibnamefont {Peper}}, \bibinfo {author} {\bibfnamefont {Y.}~\bibnamefont {Li}}, \bibinfo {author} {\bibfnamefont {D.~Y.}\ \bibnamefont {Knapp}}, \bibinfo {author} {\bibfnamefont {M.}~\bibnamefont {Bileska}}, \bibinfo {author} {\bibfnamefont {S.}~\bibnamefont {Ma}}, \bibinfo {author} {\bibfnamefont {G.}~\bibnamefont {Liu}}, \bibinfo {author} {\bibfnamefont {P.}~\bibnamefont {Peng}}, \bibinfo {author} {\bibfnamefont {B.}~\bibnamefont {Zhang}}, \bibinfo {author} {\bibfnamefont {S.~P.}\ \bibnamefont {Horvath}}, \bibinfo {author} {\bibfnamefont {A.~P.}\ \bibnamefont {Burgers}},\ and\ \bibinfo {author} {\bibfnamefont {J.~D.}\ \bibnamefont {Thompson}},\ }\bibfield  {title} {\bibinfo {title} {Spectroscopy and modeling of $^{171}\mathrm{Yb}$ rydberg states for high-fidelity two-qubit gates},\ }\href {https://doi.org/10.1103/PhysRevX.15.011009} {\bibfield  {journal} {\bibinfo  {journal} {Phys. Rev. X}\ }\textbf {\bibinfo {volume} {15}},\ \bibinfo {pages}
  {011009} (\bibinfo {year} {2025})}\BibitemShut {NoStop}%
\bibitem [{\citenamefont {Bluvstein}\ \emph {et~al.}(2024)\citenamefont {Bluvstein}, \citenamefont {Evered}, \citenamefont {Geim}, \citenamefont {Li}, \citenamefont {Zhou}, \citenamefont {Manovitz}, \citenamefont {Ebadi}, \citenamefont {Cain}, \citenamefont {Kalinowski}, \citenamefont {Hangleiter}, \citenamefont {Bonilla~Ataides}, \citenamefont {Maskara}, \citenamefont {Cong}, \citenamefont {Gao}, \citenamefont {Sales~Rodriguez}, \citenamefont {Karolyshyn}, \citenamefont {Semeghini}, \citenamefont {Gullans}, \citenamefont {Greiner}, \citenamefont {Vuleti{\'{c}}},\ and\ \citenamefont {Lukin}}]{Bluvstein2024}%
  \BibitemOpen
  \bibfield  {author} {\bibinfo {author} {\bibfnamefont {D.}~\bibnamefont {Bluvstein}}, \bibinfo {author} {\bibfnamefont {S.~J.}\ \bibnamefont {Evered}}, \bibinfo {author} {\bibfnamefont {A.~A.}\ \bibnamefont {Geim}}, \bibinfo {author} {\bibfnamefont {S.~H.}\ \bibnamefont {Li}}, \bibinfo {author} {\bibfnamefont {H.}~\bibnamefont {Zhou}}, \bibinfo {author} {\bibfnamefont {T.}~\bibnamefont {Manovitz}}, \bibinfo {author} {\bibfnamefont {S.}~\bibnamefont {Ebadi}}, \bibinfo {author} {\bibfnamefont {M.}~\bibnamefont {Cain}}, \bibinfo {author} {\bibfnamefont {M.}~\bibnamefont {Kalinowski}}, \bibinfo {author} {\bibfnamefont {D.}~\bibnamefont {Hangleiter}}, \bibinfo {author} {\bibfnamefont {J.~P.}\ \bibnamefont {Bonilla~Ataides}}, \bibinfo {author} {\bibfnamefont {N.}~\bibnamefont {Maskara}}, \bibinfo {author} {\bibfnamefont {I.}~\bibnamefont {Cong}}, \bibinfo {author} {\bibfnamefont {X.}~\bibnamefont {Gao}}, \bibinfo {author} {\bibfnamefont {P.}~\bibnamefont {Sales~Rodriguez}}, \bibinfo {author} {\bibfnamefont
  {T.}~\bibnamefont {Karolyshyn}}, \bibinfo {author} {\bibfnamefont {G.}~\bibnamefont {Semeghini}}, \bibinfo {author} {\bibfnamefont {M.~J.}\ \bibnamefont {Gullans}}, \bibinfo {author} {\bibfnamefont {M.}~\bibnamefont {Greiner}}, \bibinfo {author} {\bibfnamefont {V.}~\bibnamefont {Vuleti{\'{c}}}},\ and\ \bibinfo {author} {\bibfnamefont {M.~D.}\ \bibnamefont {Lukin}},\ }\bibfield  {title} {\bibinfo {title} {Logical quantum processor based on reconfigurable atom arrays},\ }\href {https://doi.org/10.1038/s41586-023-06927-3} {\bibfield  {journal} {\bibinfo  {journal} {Nature}\ }\textbf {\bibinfo {volume} {626}},\ \bibinfo {pages} {58} (\bibinfo {year} {2024})}\BibitemShut {NoStop}%
\bibitem [{\citenamefont {Graham}\ \emph {et~al.}(2022)\citenamefont {Graham}, \citenamefont {Song}, \citenamefont {Scott}, \citenamefont {Poole}, \citenamefont {Phuttitarn}, \citenamefont {Jooya}, \citenamefont {Eichler}, \citenamefont {Jiang}, \citenamefont {Marra}, \citenamefont {Grinkemeyer}, \citenamefont {Kwon}, \citenamefont {Ebert}, \citenamefont {Cherek}, \citenamefont {Lichtman}, \citenamefont {Gillette}, \citenamefont {Gilbert}, \citenamefont {Bowman}, \citenamefont {Ballance}, \citenamefont {Campbell}, \citenamefont {Dahl}, \citenamefont {Crawford}, \citenamefont {Blunt}, \citenamefont {Rogers}, \citenamefont {Noel},\ and\ \citenamefont {Saffman}}]{Graham2022}%
  \BibitemOpen
  \bibfield  {author} {\bibinfo {author} {\bibfnamefont {T.~M.}\ \bibnamefont {Graham}}, \bibinfo {author} {\bibfnamefont {Y.}~\bibnamefont {Song}}, \bibinfo {author} {\bibfnamefont {J.}~\bibnamefont {Scott}}, \bibinfo {author} {\bibfnamefont {C.}~\bibnamefont {Poole}}, \bibinfo {author} {\bibfnamefont {L.}~\bibnamefont {Phuttitarn}}, \bibinfo {author} {\bibfnamefont {K.}~\bibnamefont {Jooya}}, \bibinfo {author} {\bibfnamefont {P.}~\bibnamefont {Eichler}}, \bibinfo {author} {\bibfnamefont {X.}~\bibnamefont {Jiang}}, \bibinfo {author} {\bibfnamefont {A.}~\bibnamefont {Marra}}, \bibinfo {author} {\bibfnamefont {B.}~\bibnamefont {Grinkemeyer}}, \bibinfo {author} {\bibfnamefont {M.}~\bibnamefont {Kwon}}, \bibinfo {author} {\bibfnamefont {M.}~\bibnamefont {Ebert}}, \bibinfo {author} {\bibfnamefont {J.}~\bibnamefont {Cherek}}, \bibinfo {author} {\bibfnamefont {M.~T.}\ \bibnamefont {Lichtman}}, \bibinfo {author} {\bibfnamefont {M.}~\bibnamefont {Gillette}}, \bibinfo {author} {\bibfnamefont {J.}~\bibnamefont
  {Gilbert}}, \bibinfo {author} {\bibfnamefont {D.}~\bibnamefont {Bowman}}, \bibinfo {author} {\bibfnamefont {T.}~\bibnamefont {Ballance}}, \bibinfo {author} {\bibfnamefont {C.}~\bibnamefont {Campbell}}, \bibinfo {author} {\bibfnamefont {E.~D.}\ \bibnamefont {Dahl}}, \bibinfo {author} {\bibfnamefont {O.}~\bibnamefont {Crawford}}, \bibinfo {author} {\bibfnamefont {N.~S.}\ \bibnamefont {Blunt}}, \bibinfo {author} {\bibfnamefont {B.}~\bibnamefont {Rogers}}, \bibinfo {author} {\bibfnamefont {T.}~\bibnamefont {Noel}},\ and\ \bibinfo {author} {\bibfnamefont {M.}~\bibnamefont {Saffman}},\ }\bibfield  {title} {\bibinfo {title} {Multi-qubit entanglement and algorithms on a neutral-atom quantum computer},\ }\href {https://doi.org/10.1038/s41586-022-04603-6} {\bibfield  {journal} {\bibinfo  {journal} {Nature}\ }\textbf {\bibinfo {volume} {604}},\ \bibinfo {pages} {457} (\bibinfo {year} {2022})}\BibitemShut {NoStop}%
\bibitem [{\citenamefont {Ramette}\ \emph {et~al.}(2024)\citenamefont {Ramette}, \citenamefont {Sinclair}, \citenamefont {Breuckmann},\ and\ \citenamefont {Vuleti^^c4^^87}}]{Ramette2024}%
  \BibitemOpen
  \bibfield  {author} {\bibinfo {author} {\bibfnamefont {J.}~\bibnamefont {Ramette}}, \bibinfo {author} {\bibfnamefont {J.}~\bibnamefont {Sinclair}}, \bibinfo {author} {\bibfnamefont {N.~P.}\ \bibnamefont {Breuckmann}},\ and\ \bibinfo {author} {\bibfnamefont {V.}~\bibnamefont {Vuleti^^c4^^87}},\ }\bibfield  {title} {\bibinfo {title} {Fault-tolerant connection of error-corrected qubits with noisy links},\ }\href {https://doi.org/10.1038/s41534-024-00855-4} {\bibfield  {journal} {\bibinfo  {journal} {npj Quantum Information}\ }\textbf {\bibinfo {volume} {10}},\ \bibinfo {pages} {58} (\bibinfo {year} {2024})}\BibitemShut {NoStop}%
\bibitem [{\citenamefont {Pattison}\ \emph {et~al.}()\citenamefont {Pattison}, \citenamefont {Baranes}, \citenamefont {Ataides}, \citenamefont {Lukin},\ and\ \citenamefont {Zhou}}]{Pattison2024}%
  \BibitemOpen
  \bibfield  {author} {\bibinfo {author} {\bibfnamefont {C.~A.}\ \bibnamefont {Pattison}}, \bibinfo {author} {\bibfnamefont {G.}~\bibnamefont {Baranes}}, \bibinfo {author} {\bibfnamefont {J.~P.~B.}\ \bibnamefont {Ataides}}, \bibinfo {author} {\bibfnamefont {M.~D.}\ \bibnamefont {Lukin}},\ and\ \bibinfo {author} {\bibfnamefont {H.}~\bibnamefont {Zhou}},\ }\bibfield  {title} {\bibinfo {title} {Fast quantum interconnects via constant-rate entanglement distillation},\ }\href@noop {} {\ }\Eprint {https://arxiv.org/abs/2408.15936} {arXiv:2408.15936 [quant-ph]} \BibitemShut {NoStop}%
\bibitem [{\citenamefont {Li}\ and\ \citenamefont {Thompson}(2024)}]{Li2024}%
  \BibitemOpen
  \bibfield  {author} {\bibinfo {author} {\bibfnamefont {Y.}~\bibnamefont {Li}}\ and\ \bibinfo {author} {\bibfnamefont {J.~D.}\ \bibnamefont {Thompson}},\ }\bibfield  {title} {\bibinfo {title} {High-rate and high-fidelity modular interconnects between neutral atom quantum processors},\ }\href {https://doi.org/10.1103/PRXQuantum.5.020363} {\bibfield  {journal} {\bibinfo  {journal} {PRX Quantum}\ }\textbf {\bibinfo {volume} {5}},\ \bibinfo {pages} {020363} (\bibinfo {year} {2024})}\BibitemShut {NoStop}%
\bibitem [{\citenamefont {Krutyanskiy}\ \emph {et~al.}(2023)\citenamefont {Krutyanskiy}, \citenamefont {Galli}, \citenamefont {Krcmarsky}, \citenamefont {Baier}, \citenamefont {Fioretto}, \citenamefont {Pu}, \citenamefont {Mazloom}, \citenamefont {Sekatski}, \citenamefont {Canteri}, \citenamefont {Teller}, \citenamefont {Schupp}, \citenamefont {Bate}, \citenamefont {Meraner}, \citenamefont {Sangouard}, \citenamefont {Lanyon},\ and\ \citenamefont {Northup}}]{Krutyanskiy2023}%
  \BibitemOpen
  \bibfield  {author} {\bibinfo {author} {\bibfnamefont {V.}~\bibnamefont {Krutyanskiy}}, \bibinfo {author} {\bibfnamefont {M.}~\bibnamefont {Galli}}, \bibinfo {author} {\bibfnamefont {V.}~\bibnamefont {Krcmarsky}}, \bibinfo {author} {\bibfnamefont {S.}~\bibnamefont {Baier}}, \bibinfo {author} {\bibfnamefont {D.~A.}\ \bibnamefont {Fioretto}}, \bibinfo {author} {\bibfnamefont {Y.}~\bibnamefont {Pu}}, \bibinfo {author} {\bibfnamefont {A.}~\bibnamefont {Mazloom}}, \bibinfo {author} {\bibfnamefont {P.}~\bibnamefont {Sekatski}}, \bibinfo {author} {\bibfnamefont {M.}~\bibnamefont {Canteri}}, \bibinfo {author} {\bibfnamefont {M.}~\bibnamefont {Teller}}, \bibinfo {author} {\bibfnamefont {J.}~\bibnamefont {Schupp}}, \bibinfo {author} {\bibfnamefont {J.}~\bibnamefont {Bate}}, \bibinfo {author} {\bibfnamefont {M.}~\bibnamefont {Meraner}}, \bibinfo {author} {\bibfnamefont {N.}~\bibnamefont {Sangouard}}, \bibinfo {author} {\bibfnamefont {B.~P.}\ \bibnamefont {Lanyon}},\ and\ \bibinfo {author} {\bibfnamefont {T.~E.}\
  \bibnamefont {Northup}},\ }\bibfield  {title} {\bibinfo {title} {Entanglement of trapped-ion qubits separated by 230 meters},\ }\href {https://doi.org/10.1103/PhysRevLett.130.050803} {\bibfield  {journal} {\bibinfo  {journal} {Phys. Rev. Lett.}\ }\textbf {\bibinfo {volume} {130}},\ \bibinfo {pages} {050803} (\bibinfo {year} {2023})}\BibitemShut {NoStop}%
\bibitem [{\citenamefont {Hartung}\ \emph {et~al.}(2024)\citenamefont {Hartung}, \citenamefont {Seubert}, \citenamefont {Welte}, \citenamefont {Distante},\ and\ \citenamefont {Rempe}}]{Hartung2024}%
  \BibitemOpen
  \bibfield  {author} {\bibinfo {author} {\bibfnamefont {L.}~\bibnamefont {Hartung}}, \bibinfo {author} {\bibfnamefont {M.}~\bibnamefont {Seubert}}, \bibinfo {author} {\bibfnamefont {S.}~\bibnamefont {Welte}}, \bibinfo {author} {\bibfnamefont {E.}~\bibnamefont {Distante}},\ and\ \bibinfo {author} {\bibfnamefont {G.}~\bibnamefont {Rempe}},\ }\bibfield  {title} {\bibinfo {title} {A quantum-network register assembled with optical tweezers in an optical cavity},\ }\href {https://doi.org/10.1126/science.ado6471} {\bibfield  {journal} {\bibinfo  {journal} {Science}\ }\textbf {\bibinfo {volume} {385}},\ \bibinfo {pages} {179} (\bibinfo {year} {2024})}\BibitemShut {NoStop}%
\bibitem [{\citenamefont {Huie}\ \emph {et~al.}(2021)\citenamefont {Huie}, \citenamefont {Menon}, \citenamefont {Bernien},\ and\ \citenamefont {Covey}}]{Huie2021}%
  \BibitemOpen
  \bibfield  {author} {\bibinfo {author} {\bibfnamefont {W.}~\bibnamefont {Huie}}, \bibinfo {author} {\bibfnamefont {S.~G.}\ \bibnamefont {Menon}}, \bibinfo {author} {\bibfnamefont {H.}~\bibnamefont {Bernien}},\ and\ \bibinfo {author} {\bibfnamefont {J.~P.}\ \bibnamefont {Covey}},\ }\bibfield  {title} {\bibinfo {title} {Multiplexed telecommunication-band quantum networking with atom arrays in optical cavities},\ }\href {https://doi.org/10.1103/PhysRevResearch.3.043154} {\bibfield  {journal} {\bibinfo  {journal} {Phys. Rev. Res.}\ }\textbf {\bibinfo {volume} {3}},\ \bibinfo {pages} {043154} (\bibinfo {year} {2021})}\BibitemShut {NoStop}%
\bibitem [{\citenamefont {Lis}\ \emph {et~al.}(2023)\citenamefont {Lis}, \citenamefont {Senoo}, \citenamefont {McGrew}, \citenamefont {R\"onchen}, \citenamefont {Jenkins},\ and\ \citenamefont {Kaufman}}]{Lis2023}%
  \BibitemOpen
  \bibfield  {author} {\bibinfo {author} {\bibfnamefont {J.~W.}\ \bibnamefont {Lis}}, \bibinfo {author} {\bibfnamefont {A.}~\bibnamefont {Senoo}}, \bibinfo {author} {\bibfnamefont {W.~F.}\ \bibnamefont {McGrew}}, \bibinfo {author} {\bibfnamefont {F.}~\bibnamefont {R\"onchen}}, \bibinfo {author} {\bibfnamefont {A.}~\bibnamefont {Jenkins}},\ and\ \bibinfo {author} {\bibfnamefont {A.~M.}\ \bibnamefont {Kaufman}},\ }\bibfield  {title} {\bibinfo {title} {Midcircuit operations using the omg architecture in neutral atom arrays},\ }\href {https://doi.org/10.1103/PhysRevX.13.041035} {\bibfield  {journal} {\bibinfo  {journal} {Phys. Rev. X}\ }\textbf {\bibinfo {volume} {13}},\ \bibinfo {pages} {041035} (\bibinfo {year} {2023})}\BibitemShut {NoStop}%
\bibitem [{\citenamefont {Zhang}\ \emph {et~al.}()\citenamefont {Zhang}, \citenamefont {Damme}, \citenamefont {Rossignolo}, \citenamefont {Festa}, \citenamefont {Melchner}, \citenamefont {Eberhard}, \citenamefont {Tsevas}, \citenamefont {Mours}, \citenamefont {Reches}, \citenamefont {Zeiher}, \citenamefont {Blatt}, \citenamefont {Bloch}, \citenamefont {Glaser},\ and\ \citenamefont {Alberti}}]{Zhang2024}%
  \BibitemOpen
  \bibfield  {author} {\bibinfo {author} {\bibfnamefont {Z.}~\bibnamefont {Zhang}}, \bibinfo {author} {\bibfnamefont {L.~V.}\ \bibnamefont {Damme}}, \bibinfo {author} {\bibfnamefont {M.}~\bibnamefont {Rossignolo}}, \bibinfo {author} {\bibfnamefont {L.}~\bibnamefont {Festa}}, \bibinfo {author} {\bibfnamefont {M.}~\bibnamefont {Melchner}}, \bibinfo {author} {\bibfnamefont {R.}~\bibnamefont {Eberhard}}, \bibinfo {author} {\bibfnamefont {D.}~\bibnamefont {Tsevas}}, \bibinfo {author} {\bibfnamefont {K.}~\bibnamefont {Mours}}, \bibinfo {author} {\bibfnamefont {E.}~\bibnamefont {Reches}}, \bibinfo {author} {\bibfnamefont {J.}~\bibnamefont {Zeiher}}, \bibinfo {author} {\bibfnamefont {S.}~\bibnamefont {Blatt}}, \bibinfo {author} {\bibfnamefont {I.}~\bibnamefont {Bloch}}, \bibinfo {author} {\bibfnamefont {S.~J.}\ \bibnamefont {Glaser}},\ and\ \bibinfo {author} {\bibfnamefont {A.}~\bibnamefont {Alberti}},\ }\href@noop {} {\bibinfo {title} {Recoil-free quantum gates with optical qubits}},\ \Eprint
  {https://arxiv.org/abs/2408.04622} {arXiv:2408.04622 [quant-ph]} \BibitemShut {NoStop}%
\bibitem [{\citenamefont {Reiserer}\ and\ \citenamefont {Rempe}(2015)}]{Reiserer2015}%
  \BibitemOpen
  \bibfield  {author} {\bibinfo {author} {\bibfnamefont {A.}~\bibnamefont {Reiserer}}\ and\ \bibinfo {author} {\bibfnamefont {G.}~\bibnamefont {Rempe}},\ }\bibfield  {title} {\bibinfo {title} {Cavity-based quantum networks with single atoms and optical photons},\ }\href {https://doi.org/10.1103/RevModPhys.87.1379} {\bibfield  {journal} {\bibinfo  {journal} {Rev. Mod. Phys.}\ }\textbf {\bibinfo {volume} {87}},\ \bibinfo {pages} {1379} (\bibinfo {year} {2015})}\BibitemShut {NoStop}%
\bibitem [{\citenamefont {Simon}\ and\ \citenamefont {Irvine}(2003)}]{Simon2003}%
  \BibitemOpen
  \bibfield  {author} {\bibinfo {author} {\bibfnamefont {C.}~\bibnamefont {Simon}}\ and\ \bibinfo {author} {\bibfnamefont {W.~T.~M.}\ \bibnamefont {Irvine}},\ }\bibfield  {title} {\bibinfo {title} {Robust long-distance entanglement and a loophole-free bell test with ions and photons},\ }\href {https://doi.org/10.1103/PhysRevLett.91.110405} {\bibfield  {journal} {\bibinfo  {journal} {Phys. Rev. Lett.}\ }\textbf {\bibinfo {volume} {91}},\ \bibinfo {pages} {110405} (\bibinfo {year} {2003})}\BibitemShut {NoStop}%
\bibitem [{\citenamefont {Tanji}\ \emph {et~al.}(2024)\citenamefont {Tanji}, \citenamefont {Takahashi}, \citenamefont {Roga},\ and\ \citenamefont {Takeoka}}]{Tanji2024}%
  \BibitemOpen
  \bibfield  {author} {\bibinfo {author} {\bibfnamefont {K.}~\bibnamefont {Tanji}}, \bibinfo {author} {\bibfnamefont {H.}~\bibnamefont {Takahashi}}, \bibinfo {author} {\bibfnamefont {W.}~\bibnamefont {Roga}},\ and\ \bibinfo {author} {\bibfnamefont {M.}~\bibnamefont {Takeoka}},\ }\bibfield  {title} {\bibinfo {title} {Rate-fidelity tradeoff in cavity-based remote entanglement generation},\ }\href {https://doi.org/10.1103/PhysRevA.110.042405} {\bibfield  {journal} {\bibinfo  {journal} {Phys. Rev. A}\ }\textbf {\bibinfo {volume} {110}},\ \bibinfo {pages} {042405} (\bibinfo {year} {2024})}\BibitemShut {NoStop}%
\bibitem [{\citenamefont {Kikura}\ \emph {et~al.}(2025)\citenamefont {Kikura}, \citenamefont {Asaoka}, \citenamefont {Koashi},\ and\ \citenamefont {Tokunaga}}]{Kikura2024}%
  \BibitemOpen
  \bibfield  {author} {\bibinfo {author} {\bibfnamefont {S.}~\bibnamefont {Kikura}}, \bibinfo {author} {\bibfnamefont {R.}~\bibnamefont {Asaoka}}, \bibinfo {author} {\bibfnamefont {M.}~\bibnamefont {Koashi}},\ and\ \bibinfo {author} {\bibfnamefont {Y.}~\bibnamefont {Tokunaga}},\ }\bibfield  {title} {\bibinfo {title} {High-purity single-photon generation based on cavity {QED}},\ }\href {https://doi.org/10.1103/PhysRevResearch.7.013251} {\bibfield  {journal} {\bibinfo  {journal} {Phys. Rev. Res.}\ }\textbf {\bibinfo {volume} {7}},\ \bibinfo {pages} {013251} (\bibinfo {year} {2025})}\BibitemShut {NoStop}%
\bibitem [{\citenamefont {Poyatos}\ \emph {et~al.}(1996)\citenamefont {Poyatos}, \citenamefont {Cirac}, \citenamefont {Blatt},\ and\ \citenamefont {Zoller}}]{Poyatos1996}%
  \BibitemOpen
  \bibfield  {author} {\bibinfo {author} {\bibfnamefont {J.~F.}\ \bibnamefont {Poyatos}}, \bibinfo {author} {\bibfnamefont {J.~I.}\ \bibnamefont {Cirac}}, \bibinfo {author} {\bibfnamefont {R.}~\bibnamefont {Blatt}},\ and\ \bibinfo {author} {\bibfnamefont {P.}~\bibnamefont {Zoller}},\ }\bibfield  {title} {\bibinfo {title} {Trapped ions in the strong-excitation regime: Ion interferometry and nonclassical states},\ }\href {https://doi.org/10.1103/PhysRevA.54.1532} {\bibfield  {journal} {\bibinfo  {journal} {Phys. Rev. A}\ }\textbf {\bibinfo {volume} {54}},\ \bibinfo {pages} {1532} (\bibinfo {year} {1996})}\BibitemShut {NoStop}%
\bibitem [{\citenamefont {Goto}\ \emph {et~al.}(2019)\citenamefont {Goto}, \citenamefont {Mizukami}, \citenamefont {Tokunaga},\ and\ \citenamefont {Aoki}}]{Goto2019}%
  \BibitemOpen
  \bibfield  {author} {\bibinfo {author} {\bibfnamefont {H.}~\bibnamefont {Goto}}, \bibinfo {author} {\bibfnamefont {S.}~\bibnamefont {Mizukami}}, \bibinfo {author} {\bibfnamefont {Y.}~\bibnamefont {Tokunaga}},\ and\ \bibinfo {author} {\bibfnamefont {T.}~\bibnamefont {Aoki}},\ }\bibfield  {title} {\bibinfo {title} {Figure of merit for single-photon generation based on cavity quantum electrodynamics},\ }\href {https://doi.org/10.1103/PhysRevA.99.053843} {\bibfield  {journal} {\bibinfo  {journal} {Phys. Rev. A}\ }\textbf {\bibinfo {volume} {99}},\ \bibinfo {pages} {053843} (\bibinfo {year} {2019})}\BibitemShut {NoStop}%
\bibitem [{\citenamefont {Knaut}\ \emph {et~al.}(2024)\citenamefont {Knaut}, \citenamefont {Suleymanzade}, \citenamefont {Wei}, \citenamefont {Assumpcao}, \citenamefont {Stas}, \citenamefont {Huan}, \citenamefont {Machielse}, \citenamefont {Knall}, \citenamefont {Sutula}, \citenamefont {Baranes}, \citenamefont {Sinclair}, \citenamefont {De-Eknamkul}, \citenamefont {Levonian}, \citenamefont {Bhaskar}, \citenamefont {Park}, \citenamefont {Lon{\v{c}}ar},\ and\ \citenamefont {Lukin}}]{Knaut2024}%
  \BibitemOpen
  \bibfield  {author} {\bibinfo {author} {\bibfnamefont {C.~M.}\ \bibnamefont {Knaut}}, \bibinfo {author} {\bibfnamefont {A.}~\bibnamefont {Suleymanzade}}, \bibinfo {author} {\bibfnamefont {Y.-C.}\ \bibnamefont {Wei}}, \bibinfo {author} {\bibfnamefont {D.~R.}\ \bibnamefont {Assumpcao}}, \bibinfo {author} {\bibfnamefont {P.-J.}\ \bibnamefont {Stas}}, \bibinfo {author} {\bibfnamefont {Y.~Q.}\ \bibnamefont {Huan}}, \bibinfo {author} {\bibfnamefont {B.}~\bibnamefont {Machielse}}, \bibinfo {author} {\bibfnamefont {E.~N.}\ \bibnamefont {Knall}}, \bibinfo {author} {\bibfnamefont {M.}~\bibnamefont {Sutula}}, \bibinfo {author} {\bibfnamefont {G.}~\bibnamefont {Baranes}}, \bibinfo {author} {\bibfnamefont {N.}~\bibnamefont {Sinclair}}, \bibinfo {author} {\bibfnamefont {C.}~\bibnamefont {De-Eknamkul}}, \bibinfo {author} {\bibfnamefont {D.~S.}\ \bibnamefont {Levonian}}, \bibinfo {author} {\bibfnamefont {M.~K.}\ \bibnamefont {Bhaskar}}, \bibinfo {author} {\bibfnamefont {H.}~\bibnamefont {Park}}, \bibinfo {author}
  {\bibfnamefont {M.}~\bibnamefont {Lon{\v{c}}ar}},\ and\ \bibinfo {author} {\bibfnamefont {M.~D.}\ \bibnamefont {Lukin}},\ }\bibfield  {title} {\bibinfo {title} {Entanglement of nanophotonic quantum memory nodes in a telecom network},\ }\href {https://doi.org/10.1038/s41586-024-07252-z} {\bibfield  {journal} {\bibinfo  {journal} {Nature}\ }\textbf {\bibinfo {volume} {629}},\ \bibinfo {pages} {573} (\bibinfo {year} {2024})}\BibitemShut {NoStop}%
\bibitem [{\citenamefont {Kato}\ and\ \citenamefont {Aoki}(2015)}]{Kato2015}%
  \BibitemOpen
  \bibfield  {author} {\bibinfo {author} {\bibfnamefont {S.}~\bibnamefont {Kato}}\ and\ \bibinfo {author} {\bibfnamefont {T.}~\bibnamefont {Aoki}},\ }\bibfield  {title} {\bibinfo {title} {Strong coupling between a trapped single atom and an all-fiber cavity},\ }\href {https://doi.org/10.1103/PhysRevLett.115.093603} {\bibfield  {journal} {\bibinfo  {journal} {Phys. Rev. Lett.}\ }\textbf {\bibinfo {volume} {115}},\ \bibinfo {pages} {093603} (\bibinfo {year} {2015})}\BibitemShut {NoStop}%
\bibitem [{\citenamefont {Dordevi^^c4^^87}\ \emph {et~al.}(2021)\citenamefont {Dordevi^^c4^^87}, \citenamefont {Samutpraphoot}, \citenamefont {Ocola}, \citenamefont {Bernien}, \citenamefont {Grinkemeyer}, \citenamefont {Dimitrova}, \citenamefont {Vuleti^^c4^^87},\ and\ \citenamefont {Lukin}}]{Tamara2021}%
  \BibitemOpen
  \bibfield  {author} {\bibinfo {author} {\bibfnamefont {T.}~\bibnamefont {Dordevi^^c4^^87}}, \bibinfo {author} {\bibfnamefont {P.}~\bibnamefont {Samutpraphoot}}, \bibinfo {author} {\bibfnamefont {P.~L.}\ \bibnamefont {Ocola}}, \bibinfo {author} {\bibfnamefont {H.}~\bibnamefont {Bernien}}, \bibinfo {author} {\bibfnamefont {B.}~\bibnamefont {Grinkemeyer}}, \bibinfo {author} {\bibfnamefont {I.}~\bibnamefont {Dimitrova}}, \bibinfo {author} {\bibfnamefont {V.}~\bibnamefont {Vuleti^^c4^^87}},\ and\ \bibinfo {author} {\bibfnamefont {M.~D.}\ \bibnamefont {Lukin}},\ }\bibfield  {title} {\bibinfo {title} {Entanglement transport and a nanophotonic interface for atoms in optical tweezers},\ }\href {https://doi.org/10.1126/science.abi9917} {\bibfield  {journal} {\bibinfo  {journal} {Science}\ }\textbf {\bibinfo {volume} {373}},\ \bibinfo {pages} {1511} (\bibinfo {year} {2021})}\BibitemShut {NoStop}%
\bibitem [{\citenamefont {Gonz\'alez-Tudela}\ \emph {et~al.}(2024)\citenamefont {Gonz\'alez-Tudela}, \citenamefont {Reiserer}, \citenamefont {Garc\'\i{}a-Ripoll},\ and\ \citenamefont {Garc\'\i{}a-Vidal}}]{Gonzalez2024}%
  \BibitemOpen
  \bibfield  {author} {\bibinfo {author} {\bibfnamefont {A.}~\bibnamefont {Gonz\'alez-Tudela}}, \bibinfo {author} {\bibfnamefont {A.}~\bibnamefont {Reiserer}}, \bibinfo {author} {\bibfnamefont {J.~J.}\ \bibnamefont {Garc\'\i{}a-Ripoll}},\ and\ \bibinfo {author} {\bibfnamefont {F.~J.}\ \bibnamefont {Garc\'\i{}a-Vidal}},\ }\bibfield  {title} {\bibinfo {title} {{Light\textendash{}matter interactions in quantum nanophotonic devices}},\ }\href {https://doi.org/10.1038/s42254-023-00681-1} {\bibfield  {journal} {\bibinfo  {journal} {Nature Rev. Phys.}\ }\textbf {\bibinfo {volume} {6}},\ \bibinfo {pages} {166} (\bibinfo {year} {2024})}\BibitemShut {NoStop}%
\bibitem [{\citenamefont {Zheng}\ \emph {et~al.}(2022)\citenamefont {Zheng}, \citenamefont {Sharma},\ and\ \citenamefont {Borregaard}}]{Zheng2023}%
  \BibitemOpen
  \bibfield  {author} {\bibinfo {author} {\bibfnamefont {Y.}~\bibnamefont {Zheng}}, \bibinfo {author} {\bibfnamefont {H.}~\bibnamefont {Sharma}},\ and\ \bibinfo {author} {\bibfnamefont {J.}~\bibnamefont {Borregaard}},\ }\bibfield  {title} {\bibinfo {title} {Entanglement distribution with minimal memory requirements using time-bin photonic qudits},\ }\href {https://doi.org/10.1103/PRXQuantum.3.040319} {\bibfield  {journal} {\bibinfo  {journal} {PRX Quantum}\ }\textbf {\bibinfo {volume} {3}},\ \bibinfo {pages} {040319} (\bibinfo {year} {2022})}\BibitemShut {NoStop}%
\bibitem [{\citenamefont {Canteri}\ \emph {et~al.}()\citenamefont {Canteri}, \citenamefont {Koong}, \citenamefont {Bate}, \citenamefont {Winkler}, \citenamefont {Krutyanskiy},\ and\ \citenamefont {Lanyon}}]{Canteri2024}%
  \BibitemOpen
  \bibfield  {author} {\bibinfo {author} {\bibfnamefont {M.}~\bibnamefont {Canteri}}, \bibinfo {author} {\bibfnamefont {Z.~X.}\ \bibnamefont {Koong}}, \bibinfo {author} {\bibfnamefont {J.}~\bibnamefont {Bate}}, \bibinfo {author} {\bibfnamefont {A.}~\bibnamefont {Winkler}}, \bibinfo {author} {\bibfnamefont {V.}~\bibnamefont {Krutyanskiy}},\ and\ \bibinfo {author} {\bibfnamefont {B.~P.}\ \bibnamefont {Lanyon}},\ }\href@noop {} {\bibinfo {title} {A photon-interfaced ten qubit quantum network node}},\ \Eprint {https://arxiv.org/abs/2406.09480} {arXiv:2406.09480 [quant-ph]} \BibitemShut {NoStop}%
\bibitem [{\citenamefont {Apol^^c3^^adn}\ and\ \citenamefont {Nadlinger}()}]{Apolin2025}%
  \BibitemOpen
  \bibfield  {author} {\bibinfo {author} {\bibfnamefont {J.}~\bibnamefont {Apol^^c3^^adn}}\ and\ \bibinfo {author} {\bibfnamefont {D.~P.}\ \bibnamefont {Nadlinger}},\ }\href@noop {} {\bibinfo {title} {Recoil-induced errors and their correction in photon-mediated entanglement between atom qubits}},\ \Eprint {https://arxiv.org/abs/2503.16837} {arXiv:2503.16837 [quant-ph]} \BibitemShut {NoStop}%
\bibitem [{\citenamefont {Law}\ and\ \citenamefont {Kimble}(1997)}]{Law1997}%
  \BibitemOpen
  \bibfield  {author} {\bibinfo {author} {\bibfnamefont {C.~K.}\ \bibnamefont {Law}}\ and\ \bibinfo {author} {\bibfnamefont {H.~J.}\ \bibnamefont {Kimble}},\ }\bibfield  {title} {\bibinfo {title} {Deterministic generation of a bit-stream of single-photon pulses},\ }\href {https://doi.org/10.1080/09500349708231869} {\bibfield  {journal} {\bibinfo  {journal} {Journal of Modern Optics}\ }\textbf {\bibinfo {volume} {44}},\ \bibinfo {pages} {2067} (\bibinfo {year} {1997})}\BibitemShut {NoStop}%
\bibitem [{\citenamefont {Plenio}\ and\ \citenamefont {Knight}(1998)}]{Plenio1998}%
  \BibitemOpen
  \bibfield  {author} {\bibinfo {author} {\bibfnamefont {M.~B.}\ \bibnamefont {Plenio}}\ and\ \bibinfo {author} {\bibfnamefont {P.~L.}\ \bibnamefont {Knight}},\ }\bibfield  {title} {\bibinfo {title} {The quantum-jump approach to dissipative dynamics in quantum optics},\ }\href {https://doi.org/10.1103/RevModPhys.70.101} {\bibfield  {journal} {\bibinfo  {journal} {Rev. Mod. Phys.}\ }\textbf {\bibinfo {volume} {70}},\ \bibinfo {pages} {101} (\bibinfo {year} {1998})}\BibitemShut {NoStop}%
\bibitem [{\citenamefont {Johansson}\ \emph {et~al.}(2013)\citenamefont {Johansson}, \citenamefont {Nation},\ and\ \citenamefont {Nori}}]{Johansson2013}%
  \BibitemOpen
  \bibfield  {author} {\bibinfo {author} {\bibfnamefont {J.}~\bibnamefont {Johansson}}, \bibinfo {author} {\bibfnamefont {P.}~\bibnamefont {Nation}},\ and\ \bibinfo {author} {\bibfnamefont {F.}~\bibnamefont {Nori}},\ }\bibfield  {title} {\bibinfo {title} {Qutip 2: A python framework for the dynamics of open quantum systems},\ }\href {https://doi.org/https://doi.org/10.1016/j.cpc.2012.11.019} {\bibfield  {journal} {\bibinfo  {journal} {Computer Physics Communications}\ }\textbf {\bibinfo {volume} {184}},\ \bibinfo {pages} {1234} (\bibinfo {year} {2013})}\BibitemShut {NoStop}%
\bibitem [{\citenamefont {Schine}\ \emph {et~al.}(2016)\citenamefont {Schine}, \citenamefont {Ryou}, \citenamefont {Gromov}, \citenamefont {Sommer},\ and\ \citenamefont {Simon}}]{Schine2016}%
  \BibitemOpen
  \bibfield  {author} {\bibinfo {author} {\bibfnamefont {N.}~\bibnamefont {Schine}}, \bibinfo {author} {\bibfnamefont {A.}~\bibnamefont {Ryou}}, \bibinfo {author} {\bibfnamefont {A.}~\bibnamefont {Gromov}}, \bibinfo {author} {\bibfnamefont {A.}~\bibnamefont {Sommer}},\ and\ \bibinfo {author} {\bibfnamefont {J.}~\bibnamefont {Simon}},\ }\bibfield  {title} {\bibinfo {title} {Synthetic landau levels for photons},\ }\href {https://doi.org/10.1038/nature17943} {\bibfield  {journal} {\bibinfo  {journal} {Nature}\ }\textbf {\bibinfo {volume} {534}},\ \bibinfo {pages} {671} (\bibinfo {year} {2016})}\BibitemShut {NoStop}%
\bibitem [{\citenamefont {Jia}\ \emph {et~al.}(2018)\citenamefont {Jia}, \citenamefont {Schine}, \citenamefont {Georgakopoulos}, \citenamefont {Ryou}, \citenamefont {Sommer},\ and\ \citenamefont {Simon}}]{Nigyuan2018}%
  \BibitemOpen
  \bibfield  {author} {\bibinfo {author} {\bibfnamefont {N.}~\bibnamefont {Jia}}, \bibinfo {author} {\bibfnamefont {N.}~\bibnamefont {Schine}}, \bibinfo {author} {\bibfnamefont {A.}~\bibnamefont {Georgakopoulos}}, \bibinfo {author} {\bibfnamefont {A.}~\bibnamefont {Ryou}}, \bibinfo {author} {\bibfnamefont {A.}~\bibnamefont {Sommer}},\ and\ \bibinfo {author} {\bibfnamefont {J.}~\bibnamefont {Simon}},\ }\bibfield  {title} {\bibinfo {title} {Photons and polaritons in a broken-time-reversal nonplanar resonator},\ }\href {https://doi.org/10.1103/PhysRevA.97.013802} {\bibfield  {journal} {\bibinfo  {journal} {Phys. Rev. A}\ }\textbf {\bibinfo {volume} {97}},\ \bibinfo {pages} {013802} (\bibinfo {year} {2018})}\BibitemShut {NoStop}%
\bibitem [{\citenamefont {Utsugi}\ \emph {et~al.}(2022)\citenamefont {Utsugi}, \citenamefont {Goban}, \citenamefont {Tokunaga}, \citenamefont {Goto},\ and\ \citenamefont {Aoki}}]{Utsugi2022}%
  \BibitemOpen
  \bibfield  {author} {\bibinfo {author} {\bibfnamefont {T.}~\bibnamefont {Utsugi}}, \bibinfo {author} {\bibfnamefont {A.}~\bibnamefont {Goban}}, \bibinfo {author} {\bibfnamefont {Y.}~\bibnamefont {Tokunaga}}, \bibinfo {author} {\bibfnamefont {H.}~\bibnamefont {Goto}},\ and\ \bibinfo {author} {\bibfnamefont {T.}~\bibnamefont {Aoki}},\ }\bibfield  {title} {\bibinfo {title} {Gaussian-wave-packet model for single-photon generation based on cavity quantum electrodynamics under adiabatic and nonadiabatic conditions},\ }\href {https://doi.org/10.1103/PhysRevA.106.023712} {\bibfield  {journal} {\bibinfo  {journal} {Phys. Rev. A}\ }\textbf {\bibinfo {volume} {106}},\ \bibinfo {pages} {023712} (\bibinfo {year} {2022})}\BibitemShut {NoStop}%
\bibitem [{\citenamefont {Glauber}(1963)}]{Glauber1963}%
  \BibitemOpen
  \bibfield  {author} {\bibinfo {author} {\bibfnamefont {R.~J.}\ \bibnamefont {Glauber}},\ }\bibfield  {title} {\bibinfo {title} {Coherent and incoherent states of the radiation field},\ }\href {https://doi.org/10.1103/PhysRev.131.2766} {\bibfield  {journal} {\bibinfo  {journal} {Phys. Rev.}\ }\textbf {\bibinfo {volume} {131}},\ \bibinfo {pages} {2766} (\bibinfo {year} {1963})}\BibitemShut {NoStop}%
\end{thebibliography}%

\end{document}